\documentclass[aps,pra,
			   reprint,
			   a4paper,
			   longbibliography, 
			   floatfix, 
			  ]{revtex4-1}	

\usepackage{datetime} 
\usepackage[colorlinks=true,linkcolor=blue,urlcolor=blue,citecolor=blue]{hyperref}
\usepackage{graphicx}	
\usepackage[T1]{fontenc}
\usepackage{lmodern}

\RequirePackage{amsmath}
\RequirePackage{mathtools} 
\usepackage{amsfonts}
\usepackage{amssymb}
\usepackage{amsbsy}
\usepackage{dsfont}   
\usepackage{latexsym}

\usepackage[svgnames]{xcolor}
\usepackage{tikz}
\usetikzlibrary{decorations.markings}
\usetikzlibrary{shapes.geometric}
\pgfdeclarelayer{edgelayer}
\pgfdeclarelayer{nodelayer}
\pgfsetlayers{edgelayer,nodelayer,main}

\newcommand{\ket}[1]{\vert #1 \rangle}
\newcommand{\braket}[2]{\left \langle #1 \mid #2 \right \rangle}
\newcommand{\ketbra}[2]{\vert #1 \rangle \! \langle #2 \vert}

\newcommand{\sandwich}[3]{\left \langle #1 \mid #2 \mid #3 \right\rangle}

\begin{document}

\title{Practical designs for permutation symmetric problem Hamiltonians on hypercubes}
\author{A.~Ben Dodds, Viv Kendon, Charles S.~Adams, Nicholas Chancellor}
\date{\textcolor{red}{\currenttime \,\, \today}}	
\affiliation{Department of Physics, Durham University, South Road, Durham DH1~3LE, UK}

\begin{abstract}
We present a method to experimentally realize large-scale permutation-symmetric Hamiltonians for continuous-time quantum protocols such as quantum walk and adiabatic quantum computation. In particular, the method can be used to perform an encoded continuous-time quantum search on a hypercube graph with $2^n$ vertices encoded into $2n$ qubits.  We provide details for a realistically achievable implementation in Rydberg atomic systems.  Although the method is perturbative, the realization is always achieved at second order in perturbation theory, regardless of the size of the mapped system.  This highly efficient mapping provides a natural set of problems which are tractable both numerically and analytically, thereby providing a powerful tool for benchmarking quantum hardware and experimentally investigating the physics of continuous-time quantum protocols. 
\end{abstract}

\maketitle


Quantum computing based on continuous time evolution rather than discrete gate operations offers a promising route for practical near-term quantum computing. This approach has a wide variety of natural applications including in finance \cite{marzec16a,Venturelli18a,Orus18a}, aerospace \cite{coxson14a}, machine learning \cite{amin16a,Benedetti16a,Benedetti16b}, theoretical computer science \cite{chancellor16a}, mathematics \cite{Li17a,Bian13a}, decoding of communications \cite{chancellor16b} and computational biology \cite{perdomo-ortiz12a}. Moreover, experimental quantum annealing has proven highly successful recently 
\cite{brooke99a,johnson11a,denchev16a,lanting14a,boixo16a}.

While continuous-time quantum computing shows great promise, there are few known methods to experimentally implement test problems that can be used to prove the performance of hardware. For quantum computing based on discrete gates, solving unstructured search by Grover's algorithm \cite{grover97a} provides a quadratic speedup over any classical algorithm---the best possible speedup as proven by Bennett et al.~\cite{bennett97a}.  There are continuous-time variants of quantum search algorithms which can obtain the same optimal speedup for both adiabatic quantum computation \cite{farhi00a,roland02a} and continuous-time quantum walk \cite{childs03a}.  It has recently been shown that these are the two extremes of a continuum of protocols that all achieve the optimal quantum speedup \cite{morley17}. 

Continuous-time search algorithms are not easy to experimentally implement when encoded into qubits.  In contrast, Grover's original algorithm can be efficiently decomposed into quantum gates \cite{grover96a}.  A naive decomposition of the continuous-time search problem yields exponentially many terms coupling all possible subsets of qubits.  To date, the largest qubit-encoded continuous-time quantum walks have been performed on two qubits \cite{Du2003,Qiang2016}; neither implemented a search algorithm.  Larger encoded discrete-time quantum walks and quantum searches have been experimentally realized \cite{Ryan2005,Lu2010}, and alternative encodings have been explored in \cite{Matjeschk2012,Arkadiusz19a}.


Because of the difficulty of implementing qubit-encoded continuous-time quantum search algorithms, this has been considered a toy problem: useful as a theoretical tool, but not practical experimentally. 
The search Hamiltonian can always be represented in a permutation symmetric basis, by transforming the marked state to either the $\ket{0000...}$ or $\ket{11111...}$ state, although is not permutation symmetric in any other basis. However, for the purposes of this paper we are interested in the dynamics of quantum searches, which remain invariant under basis transforms, so we can consider the search problem in the symmetric basis without loss of generality.
Permutation symmetric problems have a Hamiltonian of the form
%
$H_{\mathrm{prob}}=\sum_j f\left[|j|\right] \ketbra{j}{j}$,
%
where $f$ is an arbitrary real-valued function and $|j|$ is the Hamming weight, (number of ones in $j$ when expressed as a binary number).
\begin{figure}

\tikzstyle{data}=[fill=red, draw=black, shape=circle]
\tikzstyle{auxilla}=[fill=blue, draw=black, shape=circle]
\tikzstyle{new style 0}=[fill=white, draw=black, shape=circle]

\tikzstyle{J}=[-]
\tikzstyle{Ja}=[-, draw=blue]
\begin{tikzpicture}
	\begin{pgfonlayer}{nodelayer}
		\node [style=data] (0) at (0.75, 1) {};
		\node [style=data] (1) at (0.75, -1) {};
		\node [style=data] (2) at (-1, 1) {};
		\node [style=data] (3) at (-1, -1) {};
		\node [style=auxilla] (4) at (2.5, 0) {};
		\node [style=auxilla] (5) at (1.25, 0) {};
		\node [style=auxilla] (6) at (-1.5, 0) {};
		\node [style=auxilla] (7) at (-3, 0) {};
	\end{pgfonlayer}
	\begin{pgfonlayer}{edgelayer}
		\draw [style=J] (0) to (3);
		\draw [style=J] (1) to (2);
		\draw [style=J] (2) to (0);
		\draw [style=J] (0) to (1);
		\draw [style=J] (1) to (3);
		\draw [style=J] (3) to (2);
		\draw [style=Ja] (2) to (7);
		\draw [style=Ja] (2) to (6);
		\draw [style=Ja] (0) to (5);
		\draw [style=Ja] (0) to (6);
		\draw [style=Ja] (2) to (5);
		\draw [style=Ja] (2) to (4);
		\draw [style=Ja] (0) to (7);
		\draw [style=Ja] (0) to (4);
		\draw [style=Ja] (3) to (7);
		\draw [style=Ja] (3) to (6);
		\draw [style=Ja] (3) to (5);
		\draw [style=Ja] (1) to (4);
		\draw [style=Ja] (1) to (5);
		\draw [style=Ja] (1) to (6);
		\draw [style=Ja] (1) to (7);
		\draw [style=Ja] (3) to (4);
	\end{pgfonlayer}
\end{tikzpicture}
\caption{Four qubit example of gadget coupling pattern, auxilliary qubits in blue (dark grey in print) and data qubits in red (light grey in print). Couplings corresponding to $J_a$ are blue (grey in print) and  $J$ in black. \label{fig:4local}}
\end{figure}
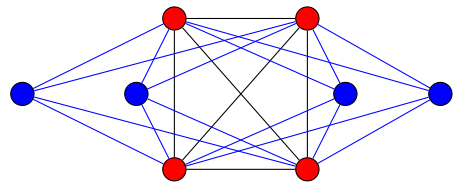
In this work, we 
present perturbative techniques for implementing permutation symmetric potentials with transverse field driving.   Importantly, the methods we present require only two-body interactions, and hence are potentially achievable in superconducting circuits and atomic systems.  Furthermore, these perturbative methods only require second order in perturbation theory, regardless of the number of qubits.  An efficient method for designing gadgets, such as the example shown in Fig.~\ref{fig:4local}, for permutation symmetric potentials has already been developed \cite{chancellor17a,chancellor16a}.  However, these papers focused only on the mapping of classical problems to quantum hardware.  To solve the problem, a driver Hamiltonian must be added, and this could spoil the performance of the gadgets.  In this work, we analyze the perturbative behavior when a transverse field driver Hamiltonian is also applied to these gadgets for permutation symmetric potentials, thereby determining their practical range of use for testbeds bench marking quantum hardware.


For a review of adiabatic quantum computation, including permutation symmetric problems, see \cite{albash16a}.
Other permutation symmetric problems include spike-like problems, first studied by Farhi et.~al.~\cite{farhi02a}, which can yield an exponential separation between the performance of adiabatic quantum computing and simulated annealing \cite{farhi02a,kong17a}. The quantum algorithm can approach a constant runtime independent of the number of qubits \cite{reichardt04a,albash16a}, while the simulated annealing runtime grows exponentially. It has been shown that the dynamics of spike-like problems can be effectively captured (at least in terms of the separation between exponential and polynomial scaling) by path integral quantum Monte Carlo, a classical simulation algorithm inspired by quantum physics \cite{Crosson14a,Brady15a,Crosson16a}. While these results make spike problems less interesting from a computational perspective, they still contain interesting many body physics, and may still provide useful tests of how faithfully the underlying quantum dynamics is reproduced in an experimental system.

Permutation symmetric plateau-like problems, have an energy landscape that becomes flat for a range of Hamming weights. For these problems, adiabatic quantum computing is polynomially faster than simulated annealing, but slower than \emph{diabatic cascades} based on rapid, non-adiabatic quenches \cite{muthukrishnan16a}. The underlying mechanism behind these cascades has been shown to be grounded in semi-classical spin mechanics, rather than fundamentally quantum behavior \cite{muthukrishnan16a}.  However, it has been demonstrated that diabatic cascades are only possible with finely tuned parameters \cite{Brady17a}.
Less work has been done on spike or plateau problems in the context of continuous-time quantum walks: this will be explored in future work \cite{fulton18a}. 

While an efficient experimental implementation of search and other problems with permutation symmetric representations could provide an effective experimental testbed, such implementations are not aimed at providing practical quantum algorithms, since the permutation symmetric problems have tractable classical algorithms.  Indeed, it is the existence of analytical and numerical solutions alongside the quantum implementation, that makes them suitable for testbed applications.  

Many techniques \cite{Dattani18a} exist to map complex classical Hamiltonians to two-body terms, (known as `quadratization').  For this work, the mapping in \cite{chancellor17a,chancellor16a} is ideal, because it can realize \emph{any} permutation symmetric problem Hamiltonian, and has a high degree of symmetry. In principle, the superconducting flux circuit construction given in \cite{chancellor17a} could be used for the gadgets proposed here.  However, in practice it is desirable to have a less noisy implementation: we therefore propose implementation in atomic systems in section \ref{sec:atoms}. Another possible implementation is a transmon interaction scheme similar to the one proposed in \cite{Leib16a}.

Along with the variety of methods which exist to exactly map classical Hamiltonians to two body terms, there are also \emph{perturbative gadgets} which are known to map higher locality quantum Hamiltonians to two-local quantum Hamiltonians perturbatively \cite{Jordan08a,Kempe04a,Biamonte07a}. While the Hamiltonian constructions we propose here can be viewed as pertubative gadgets, they differ significantly from the previous constructions and have been designed to achieve different goals. Traditionally, perturbative gadgets have focused on producing Pauli strings (e.g. $\prod_{i\in s} Z_i$) with more than two non-identity entries using only two-body terms. In these constructions, the order of perturbation theory required is equal to the number of non-identity entries in the Pauli string. Realizing the permutation symmetric problems we propose in this paper directly, using traditional perturbative gadgets, would require exponentially many such gadgets, one for each Pauli string with more than two $Z$ terms. 
Our method, on the other hand, is specifically designed to realize problems with a particular symmetry, and is realized at second order in perturbation theory regardless of the size of the problem. Our methods are therefore highly efficient special purpose perturbative gadgets, which cannot be used for everything which traditional perturbative gadgets can, but can implement a specific important class of problems.

\section{Perturbative implementation}

The gadgets proposed in \cite{chancellor17a,chancellor16a} are based on symmetric pairwise anti-ferromagnetic Ising couplings between a set of data qubits and further anti-ferromagnetic couplings between all data qubits and a set of auxiliary qubits.  This is illustrated in Fig.~\ref{fig:4local} for four data qubits. Ising field terms are applied to each of the qubits to create a low energy manifold where the total number of qubits in the logical one state $\ket{1}$ is equal to the number in the logical zero state $\ket{0}$. By placing small additional field biases on the auxilliary qubits, arbirary permutation symmetric problem Hamiltonians may be implemented in the low energy manifold. 


The gadget Hamiltonian from \cite{chancellor17a,chancellor16a} is 
\begin{eqnarray}\label{eq:Hn}
\hat{H}_n&=& J \sum_{i=1}^n \sum_{j =i+1}^{n} \hat{Z}_i \hat{Z}_j + h \sum_{i=1}^n  \hat{Z}_i \nonumber \\ 
&+& J_a \sum_{i=1}^n \sum_{j=1}^n \hat{Z}_i \hat{Z}_{j,a} +   \sum_{i=1}^n  h_{i,a} \hat{Z}_{i,a} 
\label{H2localemb}
\end{eqnarray}
acting on $n$ data qubits and $n$ auxilliary qubits, where $\hat{Z}_i$ is a Pauli-$z$ operator acting on the $i$th data qubit, and $\hat{Z}_{i,a}$ is a Pauli-$z$ operator acting on the $i$th auxilliary qubit, $J$ is the strength of symmetric two-body coupling between the data qubits, $h$ is a uniform field on the data qubits, $J_a$ is the strength of symmetric coupling between the auxilliary qubits and the data qubits, and $h_{a,i}$ is the field on the $i$th auxilliary qubit.
To realize the gadget, we set $J_a=J$, $h=-J_a+q_0$, and $h_{i,a}=-J_a(2i-n)+q_0$,
where $q_0$ is an arbitrary positive field or coupling strength. 

The form of $h_{i,a}$ ensures that the auxilliary qubits are ordered such that if one auxilliary qubit is in state $\ket{1}$, then in the low energy manifold, all auxilliary qubits with a lower index are also in state $\ket{1}$. Together, these two conditions ensure that for each state of the data qubits, there is exactly one state of the auxilliary qubits which puts the total system into the low energy manifold.  
When the Hamming weight of the data qubits is increased (decreased) by one, exactly one auxilliary qubit must be flipped from $1$ to $0$ ($0$ to $1$) to remain in the low energy manifold.
In this work, we fix $q_0=\frac{1}{2}J$, the middle of the allowed range of $q_0$ values \cite{chancellor17a,chancellor16a}.

 
We can implement a symmetric problem Hamiltonian by assigning an extra field bias exclusively to the low energy state with a particular Hamming weight. 
An energy shift of strength $2b$ can be accomplished by placing a $-b$ field on auxilliary qubit $i$ and a $+b$ field on qubit $i+1$ (or placing no field in the special case where $i=n$). 

Mathematically, we define the Hamiltonian for these extra biases $\hat{H}_{\mathrm{pot}}=\sum_{i=1}^n z_i\hat{Z}_{a,i}$ 
where, to implement a bias of strength $b_i$ on qubit $i$, we set 
\begin{equation}
z_i=\sum_{k=0}^n b_i \begin{cases}\delta_{k,i+1}-\delta_{k,i} & i \neq n \\  -\delta_{k,i} & i=n \end{cases}. \label{eq:z_pot} 
\end{equation}
%
 

%
%
Combining the two parts,
%
$\hat{H}_{\mathrm{gadg}}=\hat{H}_n+\eta \hat{H}_\mathrm{pot}$,
%
where $\hat{H}_n$ in Eq.~(\ref{eq:Hn}) creates the degenerate low energy manifold in which the auxilliary qubits count the Hamming weight of the data qubits, and  $\eta \hat{H}_\mathrm{pot}$ consists of the fields on the auxilliary qubits which create the biases that define the permutation symmetric problem. To produce a sufficiently large separation between the low energy manifold and higher energy states requires $\eta\ll1$.

\section{Transverse field driver}

We now consider what happens when we add a weak transverse field to the gadget Hamiltonian. Such transverse driving fields are usually uniform, but it will be useful to allow the transverse field strengths for the data and auxilliary qubits to be different. The Hamiltonian for the transverse fields is thus,
\begin{equation}
\hat{H}_{\mathrm{trans}}=-\gamma_d \sum_{i=1}^n \hat{X}_i-\gamma_a \sum_{i=1}^n \hat{X}_{i,a}, \label{H_trans}
\end{equation}
where $\hat{X}$ is a Pauli-$x$ operator on the specified qubit, $\gamma_d$, $\gamma_a$ set the strength of the transverse fields for the data and auxilliary qubits respectively, and the minus signs are a mathematical convenience. Setting $\gamma_d,\gamma_a\ll J$, we consider the perturbative effect of this Hamiltonian on the gadget. The action of the transverse field is to flip single qubits. Since there is no way to flip a single data or auxilliary qubit and remain in the low energy manifold, $\hat{H}_{\mathrm{trans}}$ has no effect at first order in perturbation theory. 

At second order in perturbation theory, we see that there are three possible processes which are relevant. The first process is for one data qubit to flip from $0$ to $1$ ($1$ to $0$) and an auxilliary qubit to flip from $1$ to $0$ ($0$ to $1$) in a way which leaves the final state in the low energy manifold. This process effectively maps a qubit system to the low energy manifold, with transition amplitudes proportional to $\gamma_d \, \gamma_a$. In the second process, a qubit can be flipped twice, returning to the same state, this will lead to fluctuation corrections to the energy. These corrections themselves will be permutation symmetric, so they can be corrected by applying appropriate bias fields to the auxilliary qubits. The third process is for one data qubit to flip from $1$ to $0$ and another to flip from $0$ to $1$ leaving the Hamming weight of the data qubits unchanged. Since the amplitude for this process is proportional to $\gamma_d^2$, it can be suppressed by making $\gamma_d\ll\gamma_a$.  
\begin{figure}
\includegraphics[width=4.5cm]{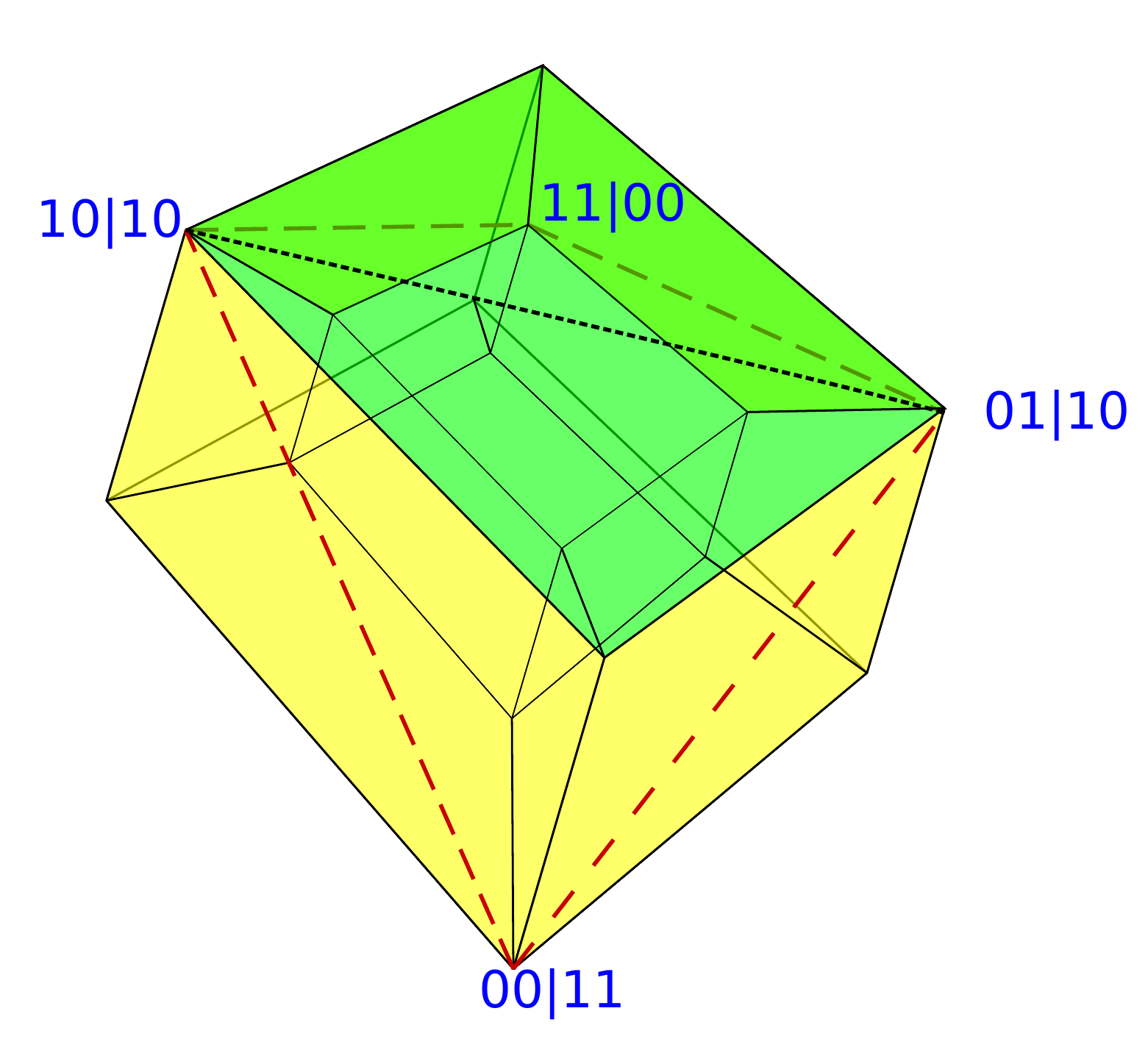} \includegraphics[width=5cm]{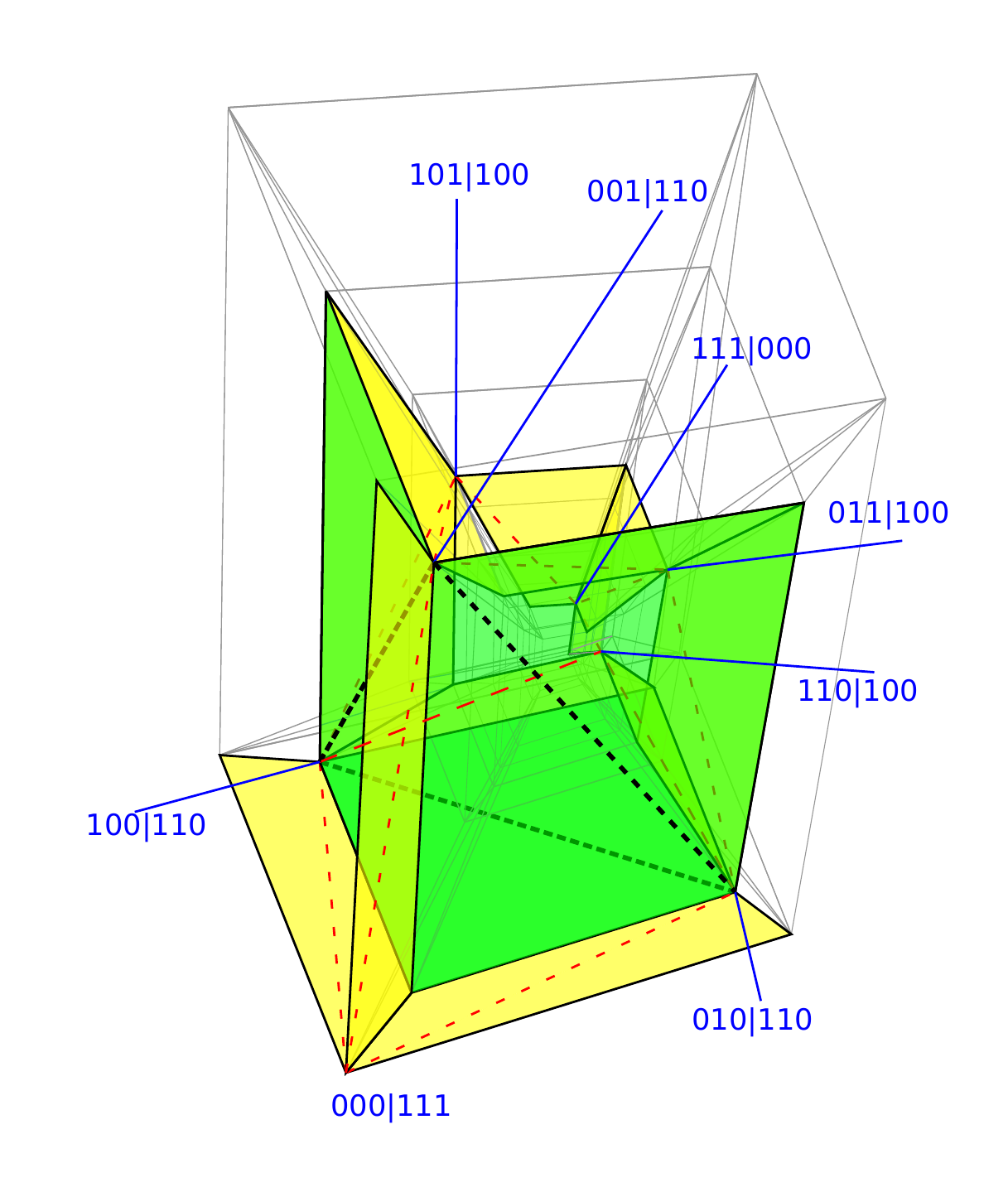}
\caption{Action of the perturbative Hamiltonians on the total solution spaces for four qubits (top) and six qubits (bottom). States are labelled as \textcolor{blue}{$\text{data qubits}| \text{ auxilliary qubits}$}. Transitions between states with different logical Hamming weights are drawn as dashed red lines, with the traversed faces colored yellow. Transitions between states with the same logical Hamming weight are drawn as black dashed lines (closer dashes) with the traversed faces colored green. The transitions between logical Hamming weight two states have been omitted from the bottom figure for visual clarity.  \label{fig:hypercube_embeddings}}
\end{figure}

The state space of the qubits forms a hypercube of dimension $2n$.  
Transitions at second order in perturbation theory correspond to diagonally traversing a face of this hypercubic. Effectively, this perturbative mapping is embedding a hypercube (the space described by the low energy manifold) onto the the faces of a hypercube of twice the dimension (the total state space of the all of the qubits). Fig.~\ref{fig:hypercube_embeddings} depicts a two dimensional projection of this embedding for two (top) and three (bottom) data qubits. For two data qubits, a two dimensional hypercube (a square) is embedded in a four dimensional hypercube (a tesseract), the square is formed by the dashed lines which connect $\textcolor{blue}{00|11}$ to $\textcolor{blue}{10|10}$ to $\textcolor{blue}{11|00}$ to $\textcolor{blue}{01|10}$ and finally back to $\textcolor{blue}{00|11}$. For three data qubits, a three dimensional hypercube (a cube) is embedded in a six dimensional hypercube. Transitions between qubits with the logical Hamming weight one are also depicted in this figure, for two data qubits, a $J(2,1)$ Johnson graph (the line segment connecting $\textcolor{blue}{10|10}$ and $\textcolor{blue}{01|10}$) is embedded in the hypercube, and for three data qubits, a $J(3,1)$ Johnson graph (a triangle) is embedded in the hypercube.


To define the perturbative mapping mathematically, we construct an effective Hamiltonian which describes the action of the Hamiltonian at second order in perturbation theory. We apply the standard textbook definition of second order perturbation theory (see e.g., \cite{LeBellac}) as well as considering the classical energies of each state in the low energy manifold.  Working in the computational basis, with $\ket{r}$, $\ket{s}$ basis states in the low energy manifold,
the effective Hamiltonian is,
\begin{align}
\sandwich{r}{\hat{H}_{\mathrm{eff}}}{s} &= \sum_q \frac{\sandwich{r}{\hat{H}_{\mathrm{trans}}}{q}\sandwich{q}{\hat{H}_{\mathrm{trans}}}{s}}{E_r-E_q}\nonumber \\  
&+ E_r \delta_{|\braket{r}{s}|,1}+O(\gamma_{a|d}\gamma_{a|d} \eta),\label{eq:2ndorderpert}
\end{align}
where $E_r$ is the energy of state $\ket{r}$ and the sum over $q$ is over all computational basis states. To determine this effective Hamiltonian, we consider different possible cases for $\ket{r}$, $\ket{s}$. We define the \emph{logical Hamming weight} $|r|$, which is the number of data qubits in the one state, and the \emph{logical Hamming distance} $\mathfrak{D}(r,s)$, which is the number of data qubits on which $\ket{r}$ and $\ket{s}$ differ. The first case we consider is $\mathfrak{D}(r,s)>2$.  In this case there is no way to transform between $\ket{r}$ and $\ket{s}$ by only two bit flips and therefore $\sandwich{r}{\hat{H}_{\mathrm{eff}}}{s}=0$. 
The next case we consider is $\mathfrak{D}(r,s)=1$ and $|r|-|s|=1$.  In this case there will always be exactly two sets of flips to go between $\ket{r}$ and $\ket{s}$, either to first flip a data qubit and then flip an auxilliary qubit, or to flip the auxilliary qubit first and then the data qubit. Geometrically, these correspond to the two ways to get from one corner of a square face of the hypercube to the other. From the form of the gadget Hamiltonian given in Eq.~(\ref{H2localemb}) there are two possibilities for the intermediate energies.  Either the intermediate energy will be $E_r-E_q=-J+O(\eta)$, if one more of the data qubits is in the $0$ state than the auxilliary qubits indicate, or $E_r-E_q=-3J+O(\eta)$ if one too many is in the $1$ state. Forturnately, for every transition there is one path through each energy manifold 
\begin{equation}
\sandwich{r}{\hat{H}_{\mathrm{eff}}}{s}=-4\frac{\gamma_d\gamma_a}{3J}+O(\gamma_d\gamma_a\eta). 
\end{equation}
These are the terms which form a hypercube on the data space. By construction, the transition terms work out to all be the same to leading order, since the strength of the penalty only depends on the number by which the auxilliary qubits `miscount', rather than the count itself.

Next, we consider the case $\mathfrak{D}(\ket{r},\ket{s})=2$ and $|(|r|-|s|)|=0$.  There will again be two possible ways to transform between the two states, corresponding to the order of the qubit flips.  In this case 
\begin{equation}
\sandwich{r}{\hat{H}_{\mathrm{eff}}}{s}=-4\frac{\gamma^2_d}{3J}+O(\gamma_d\gamma_a\eta).
\end{equation}
The final case we need to consider is $\mathfrak{D}(r,s)=0$, when $\ket{r}=\ket{s}$, which corresponds to fluctuation corrections to the energy. Because these fluctuations correspond to flipping any of the $2 n$ qubits and then flipping the same qubit back, these will not reduce to one or two simple terms. Fortunately, due to symmetry, they will be the same for states with the same logical Hamming weight. Subsituting in Eq.~(\ref{eq:2ndorderpert}) we define fluctuation terms
\begin{align}\label{eq:frakF}
& \mathcal{F}[|r|]=\sandwich{r}{\hat{H}_{\mathrm{eff}}}{r}-E_r= \\
& \sum_q \frac{\sandwich{r'}{\hat{H}_{\mathrm{trans}}}{q}\sandwich{q}{\hat{H}_{\mathrm{trans}}}{r'}}{E_{r'}-E_d}+O(\gamma_{d|a}\gamma_{d|a} \eta),\nonumber
\end{align}
where $\ket{r'}$ is a logical state where the first $|r|$ data qubits are one, and the rest are zero. This definition in terms of $\ket{r'}$ is chosen to emphasize the symmetry between states with the same Hamming weight. Combining all of these terms, we obtain the following formula for the matrix elements of $\hat{H}_{\mathrm{eff}}$.
\begin{align}
&\sandwich{r}{\hat{H}_{\mathrm{eff}}}{s}=  \nonumber \\
& \begin{cases} -4\frac{\gamma_d\gamma_a}{3J} & \mathfrak{D}(r,s)=1, |r|-|s|=\pm 1 \\ 
 -4\frac{\gamma^2_d}{3J} &\mathfrak{D}(r,s)=2, |r|-|s|=0 \\
 \mathcal{F}[|r|] +E_r & r=s\\
 0 & \mathrm{otherwise}
 \end{cases} \nonumber \\
 &+O(\gamma_x\gamma_y \eta). \label{eq:H_cases}
\end{align}

This effective Hamiltonian contains two types of unwanted terms, the fluctuation terms just analyzed, and terms which cause transitions which preserve the logical Hamming weight. The latter can be suppressed by choosing $\gamma_d \ll \gamma_a$, while the fluctuations lead to a permutation symmetric energy shift which is analytically tractable. To eliminate the effect of the fluctuations, we add $\hat{H}_{\mathrm{corr}}$, an additional bias on the auxilliary qubits. The total Hamiltonian for simulating the permutation symmetric gadget is thus
$\hat{H}_{\mathrm{sym}}=\hat{H}_n+\hat{H}_{\mathrm{trans}}+\hat{H}_{\mathrm{corr}}+\eta\hat{H}_{\mathrm{pot}}
$. 
For $\gamma_d \ll \gamma_a$, we have
$\hat{H}_{\mathrm{corr}}=\sum_{i=1}^n z_i\hat{Z}_{a,i}$, where
\begin{equation}
z_i=-\sum_{k=0}^n \mathcal{F}[k]\begin{cases}\delta_{k,i+1}-\delta_{k,i} & i \neq n \\  -\delta_{k,i} & i=n \end{cases}, \label{eq:z_gg}
\end{equation}
and $\mathcal{F}[k]$ is given by Eq.~(\ref{eq:frakF}).

If instead we have $\gamma_d \approx \gamma_a$, terms which hop between states of the same logical Hamming weight cannot be ignored. Since $J$ is positive and the Hamiltonian is permutation symmetric, the additional Hamiltonian terms which these create for each Hamming weight must have as ground states the so called Dicke states, defined as 
\begin{equation}
\ket{D_{n,k}}=\frac{1}{\sqrt{\binom{n}{k}}}\sum_{ |r|=k}\ket{r},
\end{equation}
where $\{\ket{r}\}$ are the set of states in the low energy manifold and $n$ is the total number of qubits.  Due to their symmetry, closed quantum systems initialized in a permutation symmetric state will remain in the manifold of Dicke states for all time. Therefore, if decoherence is negligible, the hopping terms between states of the same Hamming weight can be compensated by appropriately modifying  $\hat{H}_{\mathrm{corr}}$ to compensate for the additional energy shifts on the Dicke states which these terms introduce. The correction terms then take the form
\begin{equation}
z_i=-\sum_{k=0}^n [\mathcal{F}[k]+4\frac{\gamma_d^2}{3J}k(n-k)]\begin{cases}\delta_{k,i+1}-\delta_{k,i} & i \neq n \\  -\delta_{k,i} & i=n \end{cases}. \label{eq:z_apEq}
\end{equation}
Geometrically, the extra terms terms correspond to hopping on Johnson graphs embedded on the faces of the hypercube, as depicted for the four and six dimensional hypercubes in Fig.~\ref{fig:hypercube_embeddings}. If decoherence  plays a significant role in the dynamics, then states outside of the manifold of Dicke states may be populated, hence $\gamma_d \ll \gamma_a$ is required
for the mapping to be reliable with decoherence.

An astute reader may be concerned that the addition of the correction Hamiltonian $\hat{H}_{\mathrm{corr}}$ could change the original perturbative analysis and make the original assumptions no longer valid. However, this is not a concern since all of the correction terms are proportional to $\frac{\gamma^2_d}{J}$ or $\frac{\gamma_d \gamma_a}{J}$, therefore, further corrections due to shifts caused by $\hat{H}_{\mathrm{corr}}$ will be of the order $(\frac{\gamma_d}{J})^2$ and therefore small compared to $\hat{H}_{\mathrm{eff}}$, which is composed of terms of order $\frac{\gamma_d}{J}$. 

\section{Numerical validation}

Now that we have explained the mathematics behind our perturbative encoding, it remains to numerically determine the parameter values for which these gadgets work well in practice. 
\begin{figure}
\includegraphics[width=7cm]{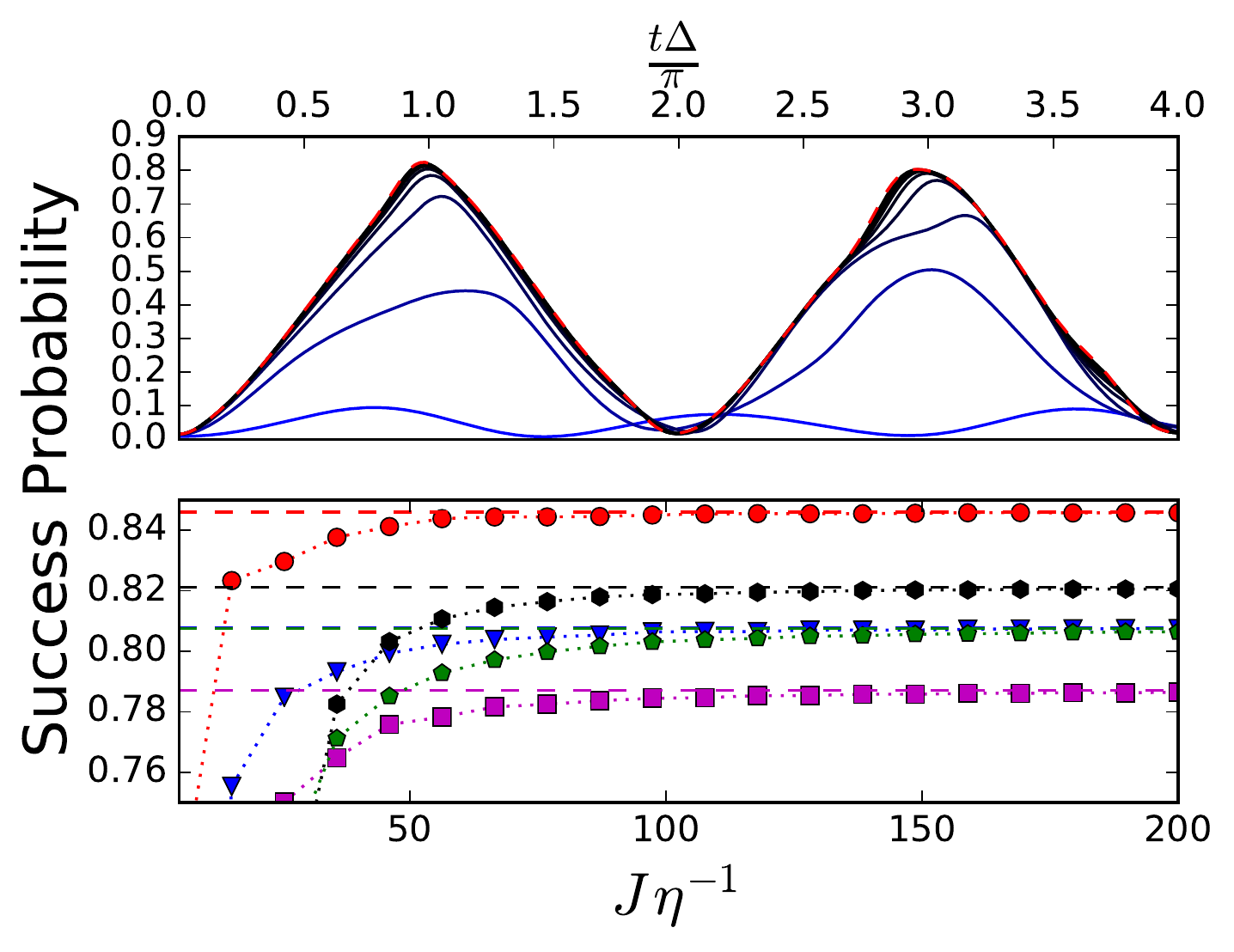}
\caption{ Top: Success probability versus time rescaled by the gap for a quantum walk on a gadget which realizes a $6$ qubit quantum search on a hypercube. $J\,\eta^{-1}$ spans $20$ linearly spaced values from $5$ (blue) to $100$ (black); red dashes the exact search Hamiltonian. Bottom: Success probability at $t=\frac{\Delta}{\pi}$ for different values of $J\,\eta^{-1}$ and different numbers of qubits, circles (red) two qubits, triangles (blue) three, squares (magenta) four, pentagons (green) five, and hexagons (black) six. Corresponding dashed lines the exact Hamiltonian.  \label{fig:search_combined}}
\end{figure}
Figure~\ref{fig:search_combined} (top) shows the results of an encoded quantum walk search on $6$ data qubits for different values of $\eta$. To achieve a search Hamiltonian, we set $\mathcal{H_{\mathrm{pot}}}=-\hat{Z}_{1,a}+\hat{Z}_{2,a}$. For this example, we choose $J=\gamma'$, $\gamma_d = \gamma_a=\eta \gamma'$ and $q_0=\frac{1}{2} \gamma'$, where
\begin{equation}
\gamma'=\gamma\,\frac{\sqrt{n}}{\left\vert\sandwich{D_{n,0}}{\hat{H}_{\mathrm{eff}}}{D_{n,1}}\right\vert},
\end{equation}
and $\gamma$ is the optimal value for quantum walk search on a hypercube \cite{childs03a,morley17}. Based on these parameter settings, $\hat{H}_{\mathrm{corr}}$ is uniquely defined by Eq.~(\ref{eq:H_cases}). We scale the runtime by $\Delta$, the gap between the ground and first excited state, thus allowing comparisons on the same plot for different values of $\eta$. We compare these results to the behavior of the exact search Hamiltonian, and see that as $\eta\rightarrow 0$, the dynamics approach those of an ideal system, as predicted by perturbation theory. If $\eta$ is chosen to be too large, then the performance is degraded as the system no longer faithfully reproduces the permutation symmetric system. 

Fig.~\ref{fig:search_combined} (bottom) shows the peak ($t=\frac{\pi}{\Delta}$) success probability of the quantum walk search versus $\eta^{-1} J$ for search gadgets of different sizes. As the value of $\eta$ becomes smaller, these peaks all approach the peak probability values obtained by the exact search Hamiltonian.

\section{Implementation}\label{sec:atoms}

In this section we propose a potential practical implementation of the scheme outlined in this paper. While there are many potential platforms which could be used to implement the necessary interactions, we have chosen Rydberg atoms \cite{Saffman16a,Sibalic18a} as they are one of the few platforms for quantum information processing that offer the flexibility of fully 3D geometries \cite{Barredo18a}. Although fidelity is lower than other systems, such as ions and superconducting qubits, recently there has been rapid progress in, for example, entanglement protocols \cite{Levine18a,Picken18a} and creating optical tweezer arrays using species such as strontium
\cite{Cooper18a,Norcia18a} and ytterbium \cite{Saskin18a}. Particularly attractive for the encoding scheme proposed here, is the possibility of all-to-all connectivity in 3D, the ability to exploit the angular dependence of the dipole-dipole couplings \cite{Saffman16a,Barredo18a}.

 While the native interactions of the Rydberg systems we consider are conditional (also known as controlled) phase shifts, not Ising interactions, controlled phase interactions can be mapped to effective Ising interactions based on the following observation. The conditional phase shift Hamiltonian shifts the phase iff both qubits are in the $\ket{1}$ state 
\begin{align}
U^{(1,2)}_{\mathrm{cond. phase}}(\phi)=\left(
\begin{array}{cccc}
1 & 0 & 0 & 0\\
0 & 1 & 0 & 0 \\
0 & 0 & 1 & 0 \\
0 & 0 & 0 & e^{-i\phi}
\end{array} \right)\nonumber \\
=\exp\left(
\begin{array}{cccc}
0 & 0 & 0 & 0\\
0 & 0 & 0 & 0 \\
0 & 0 & 0 & 0 \\
0 & 0 & 0 & -i\phi
\end{array} \right)=\exp(-i \phi \hat{C}_1\,\hat{C}_2 ),
\end{align}
where $\hat{C}_i=\frac{1}{2}(\openone-\hat{Z}_i)$.
Using some simple algebra, we observe that $\hat{Z}_i=\openone-2\,C_i$ and therefore an Ising interaction can be implemented as  $Z_i\,Z_j=4\hat{C}_i\,\hat{C}_j-2\,(\hat{C}_i+\hat{C}_j)+\openone$.
Hence, the exponentiation of the operator can be implemented by
\begin{equation}
\exp(-i \phi Z_i\,Z_j)\rightarrow U^{(i,j)}_{\mathrm{cond. phase}}(\phi)\,\exp(i\phi\,(Z_i+Z_j))
\end{equation}
up to an irrelevant global phase. This mapping corresponds exactly to the mapping of optimization problems from QUBO (quadratic unconstrained binary optimization, see: \cite{Glover18a}) form to expression as an Ising model. Translation between these two models is quite common in the context of quantum annealing \cite{Lucas14a}.

\begin{figure}\vspace*{0.5cm}
\includegraphics[width=8cm]{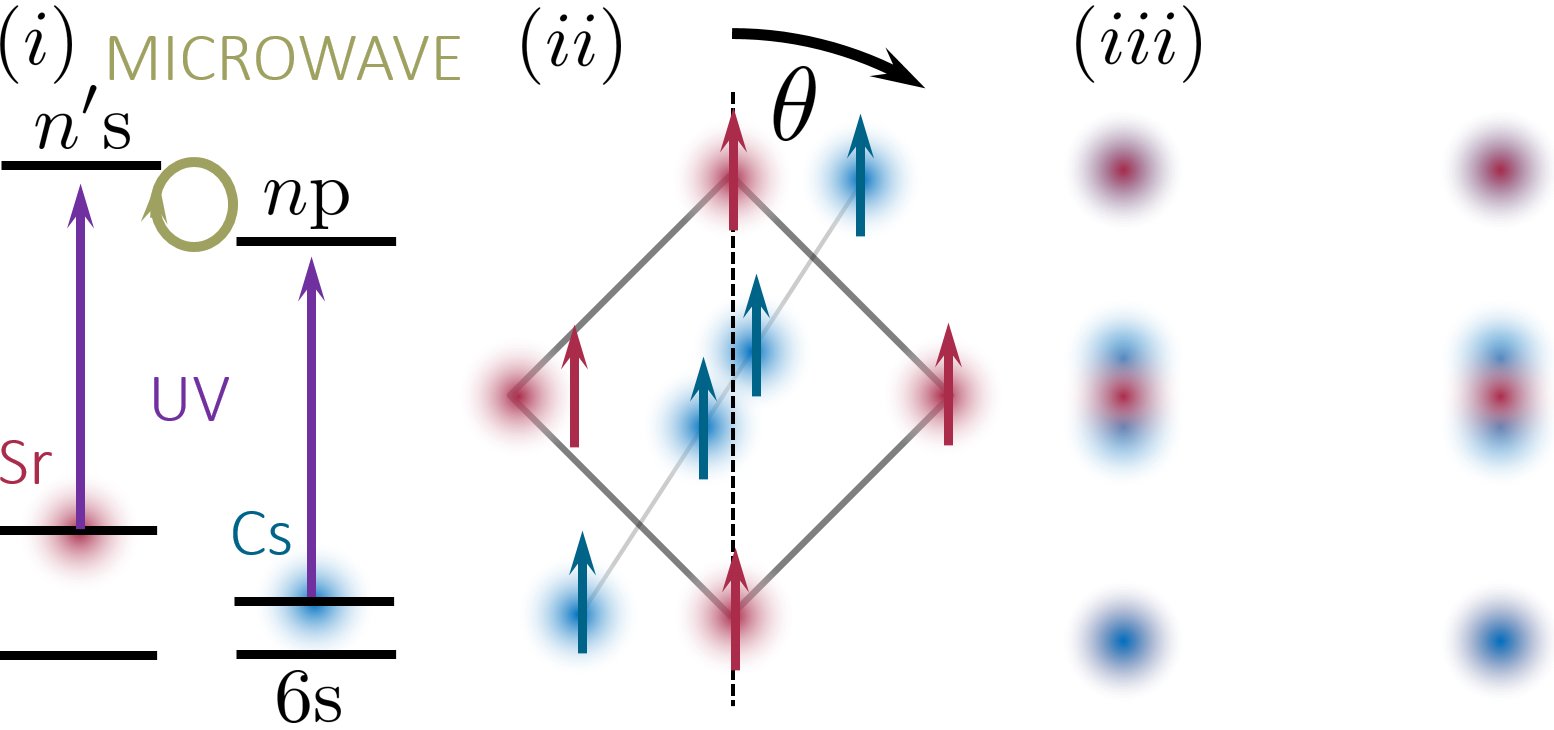}
\caption{(i) Illustration of a possible gadget level scheme. The data (red) and auxillary (blue) qubits of Fig.~1 are encoded in Sr and Cs atoms respectively. The couplings between qubits are engineered via UV excitation of Rydberg states and microwave couplings, see e.g. \cite{Paredes-Barato14a}. (ii) Frontview of the geometrical arrangement of the data and auxillary qubits. The angular dependence of $n$p Rydberg states of Cs allows us to employ a magic- angle arrangement (where $3\cos^2\theta =1$) to suppress the interaction between the auxillary qubits as required. (iii) Sideview: The protocol can be scaled either by adding additional copies of (ii) along the magic-angle diagonal or, as shown here, by adding additional planes of atoms can be added at a separation where the interaction between the auxillary qubits is insufficient to perturb the operation of the gadget.  \label{fig:magic}}
\end{figure}

Our proposed coding scheme is based on two species but could also be implemented using multiple hyperfine states in a single species. We choose cesium (Cs) and strontium (Sr), as both has been used in recent Rydberg experiments, see e.g.~\cite{Picken18a} and \cite{Bounds18a}, respectively. The data (red) and auxillary (blue) qubits of Fig.~1 are encoded in Sr and Cs atoms, respectively, as shown in Fig.~\ref{fig:magic}. The couplings between qubits are engineered via UV excitation of Rydberg states and microwave couplings between Rydberg states. The interaction between $\vert n'{\rm s}\rangle$ and $\vert n{\rm p}\rangle$ Rydberg states is of a resonant dipole-dipole type with a strength proportional to one over distance cubed, and an angular dependence $3\cos^2\theta -1$ \cite{Sibalic18a}. The interaction between atoms in the $\vert n{\rm p}\rangle$ Rydberg states also has a similar angular dependence. Therefore, if the blue atoms are positioned at the magic angle (where $3\cos^2\theta =1$) relative to the dipole axis they only interact weakly, as required for the protocol, see Fig.~\ref{fig:magic}(ii). However they still have strong interactions with the red data atoms. To scale up to large structure we can repeat the arrangement shown in Fig.~\ref{fig:magic}(ii) in adjacent places with an inter-plane distance sufficiently large that the van der Waals interactions between blue atoms in adjacent planes are below the required tolerance. All qubits not at the magic angle interact strongly, and if they are positioned within one blockade volume then it is possible to make all the interactions equal strength. While slightly more complex, and not implementing the same type of Hamiltonian, our proposed methods are very much in the spirit of the recent experimental techniques successfully demonstrated in \cite{Leseleuc18a}.
%


\section{Discussion}

Permutation symmetric problem Hamiltonians have previously been considered a theorist's tool, useful for proof-of-principle calculations, but only experimentally achievable on a very small number of qubits. In this paper, we have shown that such Hamiltonians can be realized using only one- and two-body Ising terms and a hypercube (transverse field) driver Hamiltonian, at a cost of just twice the number of qubits. Moreover, the Hamiltonian is always realized at second order in perturbation theory, regardless of size. As an example, we have outlined a way in which such gadgets can be experimentally implemented in two-species Rydberg atomic systems. 

Our work opens up the possibility of using permutation symmetric problems as testbed algorithms for benchmarking quantum computing hardware.  
Permutation symmetric problems can be understood conceptually in terms of one-dimensional potentials, and can readily be simulated numerically for thousands of qubits in the symmetric subspace. 
Note that the efficient analytical and numerical methods require knowledge of the basis in which the problem is permutation symmetric. In the unstructured search problem, for instance, this is equivalent to knowing the solution to the problem up to bit inversion. The ease of simulation does not imply that these are computationally easy problems without this extra information. 
Quantum search is extremely sensitive to the setting of the parameters \cite{childs03a,morley17}, and may be more difficult to implement experimentally than other permutation symmetric problem Hamiltonians, such as the `spike' problems discussed in \cite{farhi02a,kong17a,reichardt04a,Crosson14a,Brady15a,Crosson16a}. While spike problems cannot yield a full quantum speedup, experimental implementations could provide a powerful tool for understanding the physics of large quantum superpositions in a computational setting.  
Our perturbative gadgets provide a method to implement a wide range of permutation symmetric problems on large quantum systems, providing a powerful tool to experimentally probe the underlying physics of adiabatic quantum computing, quantum annealing, and quantum walks, as well as a method for benchmarks of, and comparisons between, different hardware.

\begin{acknowledgments}
NC and VK were funded by UK Engineering and Physical Sciences Research Council (EPSRC) grant EP/L022303/1 and NC was funded by UK EPSRC grant EP/S00114X/1. CSA was funded by UK EPSRC Grant EP/M014398. Figure \ref{fig:4local} was drawn using the Tikzit tikz editor \cite{tikzit}. We acknowledge use of Python \cite{python3},  the NumPy Python package for calculations associated with this paper \cite{numpy} , and the matplotlib Python package \cite{matplotlib} for drawing Figure \ref{fig:search_combined}. CSA thanks JD Pritchard for simulating discussions
\end{acknowledgments}


\begin{thebibliography}{64}%
\makeatletter
\providecommand \@ifxundefined [1]{%
 \@ifx{#1\undefined}
}%
\providecommand \@ifnum [1]{%
 \ifnum #1\expandafter \@firstoftwo
 \else \expandafter \@secondoftwo
 \fi
}%
\providecommand \@ifx [1]{%
 \ifx #1\expandafter \@firstoftwo
 \else \expandafter \@secondoftwo
 \fi
}%
\providecommand \natexlab [1]{#1}%
\providecommand \enquote  [1]{``#1''}%
\providecommand \bibnamefont  [1]{#1}%
\providecommand \bibfnamefont [1]{#1}%
\providecommand \citenamefont [1]{#1}%
\providecommand \href@noop [0]{\@secondoftwo}%
\providecommand \href [0]{\begingroup \@sanitize@url \@href}%
\providecommand \@href[1]{\@@startlink{#1}\@@href}%
\providecommand \@@href[1]{\endgroup#1\@@endlink}%
\providecommand \@sanitize@url [0]{\catcode `\\12\catcode `\$12\catcode
  `\&12\catcode `\#12\catcode `\^12\catcode `\_12\catcode `\%12\relax}%
\providecommand \@@startlink[1]{}%
\providecommand \@@endlink[0]{}%
\providecommand \url  [0]{\begingroup\@sanitize@url \@url }%
\providecommand \@url [1]{\endgroup\@href {#1}{\urlprefix }}%
\providecommand \urlprefix  [0]{URL }%
\providecommand \Eprint [0]{\href }%
\providecommand \doibase [0]{http://dx.doi.org/}%
\providecommand \selectlanguage [0]{\@gobble}%
\providecommand \bibinfo  [0]{\@secondoftwo}%
\providecommand \bibfield  [0]{\@secondoftwo}%
\providecommand \translation [1]{[#1]}%
\providecommand \BibitemOpen [0]{}%
\providecommand \bibitemStop [0]{}%
\providecommand \bibitemNoStop [0]{.\EOS\space}%
\providecommand \EOS [0]{\spacefactor3000\relax}%
\providecommand \BibitemShut  [1]{\csname bibitem#1\endcsname}%
\let\auto@bib@innerbib\@empty
\bibitem [{\citenamefont {Marzec}(2016)}]{marzec16a}%
  \BibitemOpen
  \bibfield  {author} {\bibinfo {author} {\bibfnamefont {Michael}\ \bibnamefont
  {Marzec}},\ }\enquote {\bibinfo {title} {Portfolio optimization: Applications
  in quantum computing},}\ in\ \href {\doibase 10.1002/9781118593486.ch4}
  {\emph {\bibinfo {booktitle} {Handbook of High-Frequency Trading and Modeling
  in Finance}}}\ (\bibinfo  {publisher} {John Wiley \& Sons, Inc.},\ \bibinfo
  {year} {2016})\ pp.\ \bibinfo {pages} {73--106}\BibitemShut {NoStop}%
\bibitem [{\citenamefont {Venturelli}\ and\ \citenamefont
  {Kondratyev}(2018)}]{Venturelli18a}%
  \BibitemOpen
  \bibfield  {author} {\bibinfo {author} {\bibfnamefont {Davide}\ \bibnamefont
  {Venturelli}}\ and\ \bibinfo {author} {\bibfnamefont {Alexei}\ \bibnamefont
  {Kondratyev}},\ }\href@noop {} {\enquote {\bibinfo {title} {Reverse quantum
  annealing approach to portfolio optimization problems},}\ } (\bibinfo {year}
  {2018}),\ \Eprint {http://arxiv.org/abs/ar$\chi$iv:1810.08584}
  {ar$\chi$iv:1810.08584} \BibitemShut {NoStop}%
\bibitem [{\citenamefont {Or{\'u}s}\ \emph {et~al.}(2018)\citenamefont
  {Or{\'u}s}, \citenamefont {Mugel},\ and\ \citenamefont {Lizaso}}]{Orus18a}%
  \BibitemOpen
  \bibfield  {author} {\bibinfo {author} {\bibfnamefont {Rom{\'a}n}\
  \bibnamefont {Or{\'u}s}}, \bibinfo {author} {\bibfnamefont {Samuel}\
  \bibnamefont {Mugel}}, \ and\ \bibinfo {author} {\bibfnamefont {Enrique}\
  \bibnamefont {Lizaso}},\ }\href@noop {} {\enquote {\bibinfo {title}
  {Forecasting financial crashes with quantum computing},}\ } (\bibinfo {year}
  {2018}),\ \Eprint {http://arxiv.org/abs/ar$\chi$iv:1810.07690}
  {ar$\chi$iv:1810.07690} \BibitemShut {NoStop}%
\bibitem [{\citenamefont {E.Coxson}\ \emph {et~al.}(2014)\citenamefont
  {E.Coxson}, \citenamefont {Hill},\ and\ \citenamefont {Russo}}]{coxson14a}%
  \BibitemOpen
  \bibfield  {author} {\bibinfo {author} {\bibfnamefont {G.}~\bibnamefont
  {E.Coxson}}, \bibinfo {author} {\bibfnamefont {C.~R.}\ \bibnamefont {Hill}},
  \ and\ \bibinfo {author} {\bibfnamefont {J.~C.}\ \bibnamefont {Russo}},\
  }\href@noop {} {\enquote {\bibinfo {title} {Adiabatic quantum computing for
  finding low-peak-sidelobe codes},}\ } (\bibinfo {year} {2014}),\ \bibinfo
  {note} {presented at the 2014 IEEE High Performance Extreme Computing
  conference}\BibitemShut {NoStop}%
\bibitem [{\citenamefont {Amin}\ \emph {et~al.}(2016)\citenamefont {Amin},
  \citenamefont {Andriyash}, \citenamefont {Rolfe}, \citenamefont
  {Kulchytskyy},\ and\ \citenamefont {Melko}}]{amin16a}%
  \BibitemOpen
  \bibfield  {author} {\bibinfo {author} {\bibfnamefont {M.~H.}\ \bibnamefont
  {Amin}}, \bibinfo {author} {\bibfnamefont {E.}~\bibnamefont {Andriyash}},
  \bibinfo {author} {\bibfnamefont {J.}~\bibnamefont {Rolfe}}, \bibinfo
  {author} {\bibfnamefont {B.}~\bibnamefont {Kulchytskyy}}, \ and\ \bibinfo
  {author} {\bibfnamefont {R.}~\bibnamefont {Melko}},\ }\href@noop {} {\enquote
  {\bibinfo {title} {Quantum {B}oltzmann machine},}\ } (\bibinfo {year}
  {2016}),\ \Eprint {http://arxiv.org/abs/arXiv:quant-ph:1601.02036}
  {arXiv:quant-ph:1601.02036} \BibitemShut {NoStop}%
\bibitem [{\citenamefont {Benedetti}\ \emph
  {et~al.}(2016{\natexlab{a}})\citenamefont {Benedetti}, \citenamefont
  {Realpe-G\'omez}, \citenamefont {Biswas},\ and\ \citenamefont
  {Perdomo-Ortiz}}]{Benedetti16a}%
  \BibitemOpen
  \bibfield  {author} {\bibinfo {author} {\bibfnamefont {M.}~\bibnamefont
  {Benedetti}}, \bibinfo {author} {\bibfnamefont {J.}~\bibnamefont
  {Realpe-G\'omez}}, \bibinfo {author} {\bibfnamefont {R.}~\bibnamefont
  {Biswas}}, \ and\ \bibinfo {author} {\bibfnamefont {A.}~\bibnamefont
  {Perdomo-Ortiz}},\ }\bibfield  {title} {\enquote {\bibinfo {title}
  {Estimation of effective temperatures in quantum annealers for sampling
  applications: A case study with possible applications in deep learning},}\
  }\href {\doibase 10.1103/PhysRevA.94.022308} {\bibfield  {journal} {\bibinfo
  {journal} {Phys. Rev. A}\ }\textbf {\bibinfo {volume} {94}},\ \bibinfo
  {pages} {022308} (\bibinfo {year} {2016}{\natexlab{a}})}\BibitemShut
  {NoStop}%
\bibitem [{\citenamefont {Benedetti}\ \emph
  {et~al.}(2016{\natexlab{b}})\citenamefont {Benedetti}, \citenamefont
  {Realpe-G{\'o}mez}, \citenamefont {Biswas},\ and\ \citenamefont
  {Perdomo-Ortiz}}]{Benedetti16b}%
  \BibitemOpen
  \bibfield  {author} {\bibinfo {author} {\bibfnamefont {M.}~\bibnamefont
  {Benedetti}}, \bibinfo {author} {\bibfnamefont {J.}~\bibnamefont
  {Realpe-G{\'o}mez}}, \bibinfo {author} {\bibfnamefont {R.}~\bibnamefont
  {Biswas}}, \ and\ \bibinfo {author} {\bibfnamefont {A.}~\bibnamefont
  {Perdomo-Ortiz}},\ }\href@noop {} {\enquote {\bibinfo {title}
  {Quantum-assisted learning of graphical models with arbitrary pairwise
  connectivity},}\ } (\bibinfo {year} {2016}{\natexlab{b}}),\ \Eprint
  {http://arxiv.org/abs/arXiv:1609.02542} {arXiv:1609.02542} \BibitemShut
  {NoStop}%
\bibitem [{\citenamefont {Chancellor}\ \emph
  {et~al.}(2016{\natexlab{a}})\citenamefont {Chancellor}, \citenamefont
  {Zohren}, \citenamefont {Warburton}, \citenamefont {Benjamin},\ and\
  \citenamefont {Roberts}}]{chancellor16a}%
  \BibitemOpen
  \bibfield  {author} {\bibinfo {author} {\bibfnamefont {N.}~\bibnamefont
  {Chancellor}}, \bibinfo {author} {\bibfnamefont {S.}~\bibnamefont {Zohren}},
  \bibinfo {author} {\bibfnamefont {P.~A.}\ \bibnamefont {Warburton}}, \bibinfo
  {author} {\bibfnamefont {S.}~\bibnamefont {Benjamin}}, \ and\ \bibinfo
  {author} {\bibfnamefont {S.}~\bibnamefont {Roberts}},\ }\bibfield  {title}
  {\enquote {\bibinfo {title} {A direct mapping of {M}ax k-{SAT} and high order
  parity checks to a chimera graph},}\ }\href {\doibase 10.1038/srep37107}
  {\bibfield  {journal} {\bibinfo  {journal} {Scientific Reports}\ }\textbf
  {\bibinfo {volume} {6}} (\bibinfo {year} {2016}{\natexlab{a}}),\
  10.1038/srep37107},\ \Eprint {http://arxiv.org/abs/arXiv:1604.00651}
  {arXiv:1604.00651} \BibitemShut {NoStop}%
\bibitem [{\citenamefont {Li}\ \emph {et~al.}(2017)\citenamefont {Li},
  \citenamefont {Dattani}, \citenamefont {Chen}, \citenamefont {Liu},
  \citenamefont {Wang}, \citenamefont {Tanburn}, \citenamefont {Chen},
  \citenamefont {Peng},\ and\ \citenamefont {Du}}]{Li17a}%
  \BibitemOpen
  \bibfield  {author} {\bibinfo {author} {\bibfnamefont {Zhaokai}\ \bibnamefont
  {Li}}, \bibinfo {author} {\bibfnamefont {Nikesh~S.}\ \bibnamefont {Dattani}},
  \bibinfo {author} {\bibfnamefont {Xi}~\bibnamefont {Chen}}, \bibinfo {author}
  {\bibfnamefont {Xiaomei}\ \bibnamefont {Liu}}, \bibinfo {author}
  {\bibfnamefont {Hengyan}\ \bibnamefont {Wang}}, \bibinfo {author}
  {\bibfnamefont {Richard}\ \bibnamefont {Tanburn}}, \bibinfo {author}
  {\bibfnamefont {Hongwei}\ \bibnamefont {Chen}}, \bibinfo {author}
  {\bibfnamefont {Xinhua}\ \bibnamefont {Peng}}, \ and\ \bibinfo {author}
  {\bibfnamefont {Jiangfeng}\ \bibnamefont {Du}},\ }\href@noop {} {\enquote
  {\bibinfo {title} {High-fidelity adiabatic quantum computation using the
  intrinsic hamiltonian of a spin system: Application to the experimental
  factorization of 291311},}\ } (\bibinfo {year} {2017}),\ \Eprint
  {http://arxiv.org/abs/1706.08061} {arXiv:1706.08061} \BibitemShut {NoStop}%
\bibitem [{\citenamefont {Bian}\ \emph {et~al.}(2013)\citenamefont {Bian},
  \citenamefont {Chudak}, \citenamefont {Macready}, \citenamefont {Clark},\
  and\ \citenamefont {Gaitan}}]{Bian13a}%
  \BibitemOpen
  \bibfield  {author} {\bibinfo {author} {\bibfnamefont {Zhengbing}\
  \bibnamefont {Bian}}, \bibinfo {author} {\bibfnamefont {Fabian}\ \bibnamefont
  {Chudak}}, \bibinfo {author} {\bibfnamefont {William~G.}\ \bibnamefont
  {Macready}}, \bibinfo {author} {\bibfnamefont {Lane}\ \bibnamefont {Clark}},
  \ and\ \bibinfo {author} {\bibfnamefont {Frank}\ \bibnamefont {Gaitan}},\
  }\bibfield  {title} {\enquote {\bibinfo {title} {Experimental determination
  of ramsey numbers},}\ }\href {\doibase 10.1103/PhysRevLett.111.130505}
  {\bibfield  {journal} {\bibinfo  {journal} {Phys. Rev. Lett.}\ }\textbf
  {\bibinfo {volume} {111}},\ \bibinfo {pages} {130505} (\bibinfo {year}
  {2013})}\BibitemShut {NoStop}%
\bibitem [{\citenamefont {Chancellor}\ \emph
  {et~al.}(2016{\natexlab{b}})\citenamefont {Chancellor}, \citenamefont
  {Szoke}, \citenamefont {Vinci}, \citenamefont {Aeppli},\ and\ \citenamefont
  {Warburton}}]{chancellor16b}%
  \BibitemOpen
  \bibfield  {author} {\bibinfo {author} {\bibfnamefont {N.}~\bibnamefont
  {Chancellor}}, \bibinfo {author} {\bibfnamefont {S.}~\bibnamefont {Szoke}},
  \bibinfo {author} {\bibfnamefont {W.}~\bibnamefont {Vinci}}, \bibinfo
  {author} {\bibfnamefont {G.}~\bibnamefont {Aeppli}}, \ and\ \bibinfo {author}
  {\bibfnamefont {P.~A.}\ \bibnamefont {Warburton}},\ }\bibfield  {title}
  {\enquote {\bibinfo {title} {Maximum--entropy inference with a programmable
  annealer},}\ }\href {\doibase doi:10.1038/srep22318} {\bibfield  {journal}
  {\bibinfo  {journal} {Scientific Reports}\ }\textbf {\bibinfo {volume} {6}}
  (\bibinfo {year} {2016}{\natexlab{b}}),\ doi:10.1038/srep22318}\BibitemShut
  {NoStop}%
\bibitem [{\citenamefont {Perdomo-Ortiz}\ \emph {et~al.}(2012)\citenamefont
  {Perdomo-Ortiz}, \citenamefont {Dickson}, \citenamefont {Drew-Brook},
  \citenamefont {Rose},\ and\ \citenamefont {Aspuru-Guzik}}]{perdomo-ortiz12a}%
  \BibitemOpen
  \bibfield  {author} {\bibinfo {author} {\bibfnamefont {Alejandro}\
  \bibnamefont {Perdomo-Ortiz}}, \bibinfo {author} {\bibfnamefont {Neil}\
  \bibnamefont {Dickson}}, \bibinfo {author} {\bibfnamefont {Marshall}\
  \bibnamefont {Drew-Brook}}, \bibinfo {author} {\bibfnamefont {Geordie}\
  \bibnamefont {Rose}}, \ and\ \bibinfo {author} {\bibfnamefont {Alan}\
  \bibnamefont {Aspuru-Guzik}},\ }\bibfield  {title} {\enquote {\bibinfo
  {title} {Finding low-energy conformations of lattice protein models by
  quantum annealing},}\ }\href@noop {} {\bibfield  {journal} {\bibinfo
  {journal} {Scientific Reports}\ }\textbf {\bibinfo {volume} {2}} (\bibinfo
  {year} {2012})}\BibitemShut {NoStop}%
\bibitem [{\citenamefont {Brooke}\ \emph {et~al.}(1999)\citenamefont {Brooke},
  \citenamefont {Bitko}, \citenamefont {Rosenbaum},\ and\ \citenamefont
  {Aeppli}}]{brooke99a}%
  \BibitemOpen
  \bibfield  {author} {\bibinfo {author} {\bibfnamefont {J.}~\bibnamefont
  {Brooke}}, \bibinfo {author} {\bibfnamefont {D.}~\bibnamefont {Bitko}},
  \bibinfo {author} {\bibfnamefont {T.~F.}\ \bibnamefont {Rosenbaum}}, \ and\
  \bibinfo {author} {\bibfnamefont {G.}~\bibnamefont {Aeppli}},\ }\bibfield
  {title} {\enquote {\bibinfo {title} {Quantum annealing of a disordered
  magnet},}\ }\href {\doibase 10.1126/science.284.5415.779} {\bibfield
  {journal} {\bibinfo  {journal} {Science}\ }\textbf {\bibinfo {volume}
  {284}},\ \bibinfo {pages} {779--781} (\bibinfo {year} {1999})},\ \Eprint
  {http://arxiv.org/abs/http://science.sciencemag.org/content/284/5415/779.full.pdf}
  {http://science.sciencemag.org/content/284/5415/779.full.pdf} \BibitemShut
  {NoStop}%
\bibitem [{\citenamefont {Johnson}\ \emph {et~al.}(2011)\citenamefont
  {Johnson}, \citenamefont {Amin}, \citenamefont {Gildert}, \citenamefont
  {Lanting}, \citenamefont {Hamze}, \citenamefont {Dickson}, \citenamefont
  {Harris}, \citenamefont {Berkley}, \citenamefont {Johansson}, \citenamefont
  {Bunyk}, \citenamefont {Chapple}, \citenamefont {Enderud}, \citenamefont
  {Hilton}, \citenamefont {Karimi}, \citenamefont {Ladizinsky}, \citenamefont
  {Ladizinsky}, \citenamefont {Oh}, \citenamefont {Perminov}, \citenamefont
  {Rich}, \citenamefont {Thom}, \citenamefont {Tolkacheva}, \citenamefont
  {Truncik}, \citenamefont {Uchaikin}, \citenamefont {Wang},\ and\
  \citenamefont {Rose}}]{johnson11a}%
  \BibitemOpen
  \bibfield  {author} {\bibinfo {author} {\bibfnamefont {M.~W.}\ \bibnamefont
  {Johnson}}, \bibinfo {author} {\bibfnamefont {M.~H.~S.}\ \bibnamefont
  {Amin}}, \bibinfo {author} {\bibfnamefont {S.}~\bibnamefont {Gildert}},
  \bibinfo {author} {\bibfnamefont {T.}~\bibnamefont {Lanting}}, \bibinfo
  {author} {\bibfnamefont {F.}~\bibnamefont {Hamze}}, \bibinfo {author}
  {\bibfnamefont {N.}~\bibnamefont {Dickson}}, \bibinfo {author} {\bibfnamefont
  {R.}~\bibnamefont {Harris}}, \bibinfo {author} {\bibfnamefont {A.~J.}\
  \bibnamefont {Berkley}}, \bibinfo {author} {\bibfnamefont {J.}~\bibnamefont
  {Johansson}}, \bibinfo {author} {\bibfnamefont {P.}~\bibnamefont {Bunyk}},
  \bibinfo {author} {\bibfnamefont {E.~M.}\ \bibnamefont {Chapple}}, \bibinfo
  {author} {\bibfnamefont {C.}~\bibnamefont {Enderud}}, \bibinfo {author}
  {\bibfnamefont {J.~P.}\ \bibnamefont {Hilton}}, \bibinfo {author}
  {\bibfnamefont {K.}~\bibnamefont {Karimi}}, \bibinfo {author} {\bibfnamefont
  {E.}~\bibnamefont {Ladizinsky}}, \bibinfo {author} {\bibfnamefont
  {N.}~\bibnamefont {Ladizinsky}}, \bibinfo {author} {\bibfnamefont
  {T.}~\bibnamefont {Oh}}, \bibinfo {author} {\bibfnamefont {I.}~\bibnamefont
  {Perminov}}, \bibinfo {author} {\bibfnamefont {C.}~\bibnamefont {Rich}},
  \bibinfo {author} {\bibfnamefont {M.~C.}\ \bibnamefont {Thom}}, \bibinfo
  {author} {\bibfnamefont {E.}~\bibnamefont {Tolkacheva}}, \bibinfo {author}
  {\bibfnamefont {C.~J.~S.}\ \bibnamefont {Truncik}}, \bibinfo {author}
  {\bibfnamefont {S.}~\bibnamefont {Uchaikin}}, \bibinfo {author}
  {\bibfnamefont {J.}~\bibnamefont {Wang}}, \ and\ \bibinfo {author}
  {\bibfnamefont {B.~Wilsonand~G.}\ \bibnamefont {Rose}},\ }\bibfield  {title}
  {\enquote {\bibinfo {title} {Quantum annealing with manufactured spins},}\
  }\href {\doibase doi:10.1038/nature10012} {\bibfield  {journal} {\bibinfo
  {journal} {Nature}\ }\textbf {\bibinfo {volume} {473}},\ \bibinfo {pages}
  {194--198} (\bibinfo {year} {2011})}\BibitemShut {NoStop}%
\bibitem [{\citenamefont {Denchev}\ \emph {et~al.}(2016)\citenamefont
  {Denchev}, \citenamefont {Boixo}, \citenamefont {Isakov}, \citenamefont
  {Ding}, \citenamefont {Babbush}, \citenamefont {Smelyanskiy}, \citenamefont
  {Martinis},\ and\ \citenamefont {Neven}}]{denchev16a}%
  \BibitemOpen
  \bibfield  {author} {\bibinfo {author} {\bibfnamefont {Vasil~S.}\
  \bibnamefont {Denchev}}, \bibinfo {author} {\bibfnamefont {Sergio}\
  \bibnamefont {Boixo}}, \bibinfo {author} {\bibfnamefont {Sergei~V.}\
  \bibnamefont {Isakov}}, \bibinfo {author} {\bibfnamefont {Nan}\ \bibnamefont
  {Ding}}, \bibinfo {author} {\bibfnamefont {Ryan}\ \bibnamefont {Babbush}},
  \bibinfo {author} {\bibfnamefont {Vadim}\ \bibnamefont {Smelyanskiy}},
  \bibinfo {author} {\bibfnamefont {John}\ \bibnamefont {Martinis}}, \ and\
  \bibinfo {author} {\bibfnamefont {Hartmut}\ \bibnamefont {Neven}},\
  }\bibfield  {title} {\enquote {\bibinfo {title} {What is the computational
  value of finite-range tunneling?}}\ }\href {\doibase
  10.1103/PhysRevX.6.031015} {\bibfield  {journal} {\bibinfo  {journal} {Phys.
  Rev. X}\ }\textbf {\bibinfo {volume} {6}},\ \bibinfo {pages} {031015}
  (\bibinfo {year} {2016})}\BibitemShut {NoStop}%
\bibitem [{\citenamefont {Lanting}\ \emph {et~al.}(2014)\citenamefont
  {Lanting}, \citenamefont {Przybysz}, \citenamefont {Smirnov}, \citenamefont
  {Spedalieri}, \citenamefont {Amin}, \citenamefont {Berkley}, \citenamefont
  {Harris}, \citenamefont {Altomare}, \citenamefont {Boixo}, \citenamefont
  {Bunyk}, \citenamefont {Dickson}, \citenamefont {Enderud}, \citenamefont
  {Hilton}, \citenamefont {Hoskinson}, \citenamefont {Johnson}, \citenamefont
  {Ladizinsky}, \citenamefont {Ladizinsky}, \citenamefont {Neufeld},
  \citenamefont {Oh}, \citenamefont {Perminov}, \citenamefont {Rich},
  \citenamefont {Thom}, \citenamefont {Tolkacheva}, \citenamefont {Uchaikin},
  \citenamefont {Wilson},\ and\ \citenamefont {Rose}}]{lanting14a}%
  \BibitemOpen
  \bibfield  {author} {\bibinfo {author} {\bibfnamefont {T.}~\bibnamefont
  {Lanting}}, \bibinfo {author} {\bibfnamefont {A.~J.}\ \bibnamefont
  {Przybysz}}, \bibinfo {author} {\bibfnamefont {A.~Yu.}\ \bibnamefont
  {Smirnov}}, \bibinfo {author} {\bibfnamefont {F.~M.}\ \bibnamefont
  {Spedalieri}}, \bibinfo {author} {\bibfnamefont {M.~H.}\ \bibnamefont
  {Amin}}, \bibinfo {author} {\bibfnamefont {A.~J.}\ \bibnamefont {Berkley}},
  \bibinfo {author} {\bibfnamefont {R.}~\bibnamefont {Harris}}, \bibinfo
  {author} {\bibfnamefont {F.}~\bibnamefont {Altomare}}, \bibinfo {author}
  {\bibfnamefont {S.}~\bibnamefont {Boixo}}, \bibinfo {author} {\bibfnamefont
  {P.}~\bibnamefont {Bunyk}}, \bibinfo {author} {\bibfnamefont
  {N.}~\bibnamefont {Dickson}}, \bibinfo {author} {\bibfnamefont
  {C.}~\bibnamefont {Enderud}}, \bibinfo {author} {\bibfnamefont {J.~P.}\
  \bibnamefont {Hilton}}, \bibinfo {author} {\bibfnamefont {E.}~\bibnamefont
  {Hoskinson}}, \bibinfo {author} {\bibfnamefont {M.~W.}\ \bibnamefont
  {Johnson}}, \bibinfo {author} {\bibfnamefont {E.}~\bibnamefont {Ladizinsky}},
  \bibinfo {author} {\bibfnamefont {N.}~\bibnamefont {Ladizinsky}}, \bibinfo
  {author} {\bibfnamefont {R.}~\bibnamefont {Neufeld}}, \bibinfo {author}
  {\bibfnamefont {T.}~\bibnamefont {Oh}}, \bibinfo {author} {\bibfnamefont
  {I.}~\bibnamefont {Perminov}}, \bibinfo {author} {\bibfnamefont
  {C.}~\bibnamefont {Rich}}, \bibinfo {author} {\bibfnamefont {M.~C.}\
  \bibnamefont {Thom}}, \bibinfo {author} {\bibfnamefont {E.}~\bibnamefont
  {Tolkacheva}}, \bibinfo {author} {\bibfnamefont {S.}~\bibnamefont
  {Uchaikin}}, \bibinfo {author} {\bibfnamefont {A.~B.}\ \bibnamefont
  {Wilson}}, \ and\ \bibinfo {author} {\bibfnamefont {G.}~\bibnamefont
  {Rose}},\ }\bibfield  {title} {\enquote {\bibinfo {title} {Entanglement in a
  quantum annealing processor},}\ }\href {\doibase 10.1103/PhysRevX.4.021041}
  {\bibfield  {journal} {\bibinfo  {journal} {Phys. Rev. X}\ }\textbf {\bibinfo
  {volume} {4}},\ \bibinfo {pages} {021041} (\bibinfo {year}
  {2014})}\BibitemShut {NoStop}%
\bibitem [{\citenamefont {Boixo}\ \emph {et~al.}(2016)\citenamefont {Boixo},
  \citenamefont {Smelyanskiy}, \citenamefont {Shabani}, \citenamefont {Isakov},
  \citenamefont {Dykman}, \citenamefont {Denchev}, \citenamefont {Amin},
  \citenamefont {Smirnov}, \citenamefont {Mohseni},\ and\ \citenamefont
  {Neven}}]{boixo16a}%
  \BibitemOpen
  \bibfield  {author} {\bibinfo {author} {\bibfnamefont {Sergio}\ \bibnamefont
  {Boixo}}, \bibinfo {author} {\bibfnamefont {Vadim~N.}\ \bibnamefont
  {Smelyanskiy}}, \bibinfo {author} {\bibfnamefont {Alireza}\ \bibnamefont
  {Shabani}}, \bibinfo {author} {\bibfnamefont {Sergei~V.}\ \bibnamefont
  {Isakov}}, \bibinfo {author} {\bibfnamefont {Mark}\ \bibnamefont {Dykman}},
  \bibinfo {author} {\bibfnamefont {Vasil~S.}\ \bibnamefont {Denchev}},
  \bibinfo {author} {\bibfnamefont {Mohammad~H.}\ \bibnamefont {Amin}},
  \bibinfo {author} {\bibfnamefont {Anatoly~Yu}\ \bibnamefont {Smirnov}},
  \bibinfo {author} {\bibfnamefont {Masoud}\ \bibnamefont {Mohseni}}, \ and\
  \bibinfo {author} {\bibfnamefont {Hartmut}\ \bibnamefont {Neven}},\
  }\bibfield  {title} {\enquote {\bibinfo {title} {Computational multiqubit
  tunnelling in programmable quantum annealers},}\ }\href {\doibase
  doi:10.1038/ncomms10327} {\bibfield  {journal} {\bibinfo  {journal} {Nature
  Communications}\ }\textbf {\bibinfo {volume} {7}} (\bibinfo {year} {2016}),\
  doi:10.1038/ncomms10327}\BibitemShut {NoStop}%
\bibitem [{\citenamefont {Grover}(1997)}]{grover97a}%
  \BibitemOpen
  \bibfield  {author} {\bibinfo {author} {\bibfnamefont {L.~K.}\ \bibnamefont
  {Grover}},\ }\bibfield  {title} {\enquote {\bibinfo {title} {Quantum
  mechanics helps in searching for a needle in a haystack},}\ }\href@noop {}
  {\bibfield  {journal} {\bibinfo  {journal} {Phys.~Rev.~Lett.}\ }\textbf
  {\bibinfo {volume} {79}},\ \bibinfo {pages} {325} (\bibinfo {year} {1997})},\
  \Eprint {http://arxiv.org/abs/arXiv:quant-ph/9706033}
  {arXiv:quant-ph/9706033} \BibitemShut {NoStop}%
\bibitem [{\citenamefont {Bennett}\ \emph {et~al.}(1997)\citenamefont
  {Bennett}, \citenamefont {Bernstein}, \citenamefont {Brassard},\ and\
  \citenamefont {Vazirani}}]{bennett97a}%
  \BibitemOpen
  \bibfield  {author} {\bibinfo {author} {\bibfnamefont {Charles~H.}\
  \bibnamefont {Bennett}}, \bibinfo {author} {\bibfnamefont {Ethan}\
  \bibnamefont {Bernstein}}, \bibinfo {author} {\bibfnamefont {Gilles}\
  \bibnamefont {Brassard}}, \ and\ \bibinfo {author} {\bibfnamefont {Umesh}\
  \bibnamefont {Vazirani}},\ }\bibfield  {title} {\enquote {\bibinfo {title}
  {Strengths and weaknesses of quantum computing},}\ }\href@noop {} {\bibfield
  {journal} {\bibinfo  {journal} {SIAM J.~Comput.}\ }\textbf {\bibinfo {volume}
  {26}},\ \bibinfo {pages} {151--152} (\bibinfo {year} {1997})}\BibitemShut
  {NoStop}%
\bibitem [{\citenamefont {Farhi}\ \emph {et~al.}(2000)\citenamefont {Farhi},
  \citenamefont {Goldstone}, \citenamefont {Gutmann},\ and\ \citenamefont
  {Sipser}}]{farhi00a}%
  \BibitemOpen
  \bibfield  {author} {\bibinfo {author} {\bibfnamefont {E.}~\bibnamefont
  {Farhi}}, \bibinfo {author} {\bibfnamefont {J.}~\bibnamefont {Goldstone}},
  \bibinfo {author} {\bibfnamefont {S.}~\bibnamefont {Gutmann}}, \ and\
  \bibinfo {author} {\bibfnamefont {M.}~\bibnamefont {Sipser}},\ }\href@noop {}
  {\enquote {\bibinfo {title} {Quantum computation by adiabatic evolution},}\ }
  (\bibinfo {year} {2000}),\ \Eprint {http://arxiv.org/abs/quant-ph/0001106}
  {quant-ph/0001106} \BibitemShut {NoStop}%
\bibitem [{\citenamefont {Roland}\ and\ \citenamefont
  {Cerf}(2002)}]{roland02a}%
  \BibitemOpen
  \bibfield  {author} {\bibinfo {author} {\bibfnamefont {J\'er\'emie}\
  \bibnamefont {Roland}}\ and\ \bibinfo {author} {\bibfnamefont {Nicolas~J.}\
  \bibnamefont {Cerf}},\ }\bibfield  {title} {\enquote {\bibinfo {title}
  {Quantum search by local adiabatic evolution},}\ }\href {\doibase
  10.1103/PhysRevA.65.042308} {\bibfield  {journal} {\bibinfo  {journal}
  {Phys.~Rev.~A}\ }\textbf {\bibinfo {volume} {65}},\ \bibinfo {pages} {042308}
  (\bibinfo {year} {2002})}\BibitemShut {NoStop}%
\bibitem [{\citenamefont {Childs}\ and\ \citenamefont
  {Goldstone}(2004)}]{childs03a}%
  \BibitemOpen
  \bibfield  {author} {\bibinfo {author} {\bibfnamefont {Andrew}\ \bibnamefont
  {Childs}}\ and\ \bibinfo {author} {\bibfnamefont {Jeffrey}\ \bibnamefont
  {Goldstone}},\ }\bibfield  {title} {\enquote {\bibinfo {title} {Spatial
  search by quantum walk},}\ }\href@noop {} {\bibfield  {journal} {\bibinfo
  {journal} {Phys.~Rev.~A}\ }\textbf {\bibinfo {volume} {70}},\ \bibinfo
  {pages} {022314} (\bibinfo {year} {2004})},\ \Eprint
  {http://arxiv.org/abs/quant-ph/0306054} {quant-ph/0306054} \BibitemShut
  {NoStop}%
\bibitem [{\citenamefont {Morley}\ \emph {et~al.}(2019)\citenamefont {Morley},
  \citenamefont {Chancellor}, \citenamefont {Bose},\ and\ \citenamefont
  {Kendon}}]{morley17}%
  \BibitemOpen
  \bibfield  {author} {\bibinfo {author} {\bibfnamefont {James~G}\ \bibnamefont
  {Morley}}, \bibinfo {author} {\bibfnamefont {Nicholas}\ \bibnamefont
  {Chancellor}}, \bibinfo {author} {\bibfnamefont {Sougato}\ \bibnamefont
  {Bose}}, \ and\ \bibinfo {author} {\bibfnamefont {Viv}\ \bibnamefont
  {Kendon}},\ }\bibfield  {title} {\enquote {\bibinfo {title} {Quantum search
  with hybrid adiabatic-quantum walk algorithms and realistic noise},}\
  }\href@noop {} {\bibfield  {journal} {\bibinfo  {journal} {Phys.~Rev.~A}\ }
  (\bibinfo {year} {2019})},\ \bibinfo {note} {to appear.},\ \Eprint
  {http://arxiv.org/abs/1709.00371} {arXiv:1709.00371} \BibitemShut {NoStop}%
\bibitem [{\citenamefont {Grover}(1996)}]{grover96a}%
  \BibitemOpen
  \bibfield  {author} {\bibinfo {author} {\bibfnamefont {Lov}\ \bibnamefont
  {Grover}},\ }\bibfield  {title} {\enquote {\bibinfo {title} {A fast quantum
  mechanical algorithm for database search},}\ }\href@noop {} {\bibfield
  {journal} {\bibinfo  {journal} {Proceedings, 28th Annual ACM Symposium on the
  Theory of Computing (STOC)}\ ,\ \bibinfo {pages} {212--219}} (\bibinfo {year}
  {1996})},\ \Eprint {http://arxiv.org/abs/arXiv:quant-ph/9605043}
  {arXiv:quant-ph/9605043} \BibitemShut {NoStop}%
\bibitem [{\citenamefont {Du}\ \emph {et~al.}(2003)\citenamefont {Du},
  \citenamefont {Li}, \citenamefont {Xu}, \citenamefont {Shi}, \citenamefont
  {Wu}, \citenamefont {Zhou},\ and\ \citenamefont {Han}}]{Du2003}%
  \BibitemOpen
  \bibfield  {author} {\bibinfo {author} {\bibfnamefont {Jiangfeng}\
  \bibnamefont {Du}}, \bibinfo {author} {\bibfnamefont {Hui}\ \bibnamefont
  {Li}}, \bibinfo {author} {\bibfnamefont {Xiaodong}\ \bibnamefont {Xu}},
  \bibinfo {author} {\bibfnamefont {Mingjun}\ \bibnamefont {Shi}}, \bibinfo
  {author} {\bibfnamefont {Jihui}\ \bibnamefont {Wu}}, \bibinfo {author}
  {\bibfnamefont {Xianyi}\ \bibnamefont {Zhou}}, \ and\ \bibinfo {author}
  {\bibfnamefont {Rongdian}\ \bibnamefont {Han}},\ }\bibfield  {title}
  {\enquote {\bibinfo {title} {Experimental implementation of the quantum
  random-walk algorithm},}\ }\href {\doibase 10.1103/PhysRevA.67.042316}
  {\bibfield  {journal} {\bibinfo  {journal} {Phys. Rev. A}\ }\textbf {\bibinfo
  {volume} {67}},\ \bibinfo {pages} {042316} (\bibinfo {year}
  {2003})}\BibitemShut {NoStop}%
\bibitem [{\citenamefont {Qiang}\ \emph {et~al.}(2016)\citenamefont {Qiang},
  \citenamefont {Loke}, \citenamefont {Montanaro}, \citenamefont
  {Aungskunsiri}, \citenamefont {Zhou}, \citenamefont {O'Brien}, \citenamefont
  {Wang},\ and\ \citenamefont {Matthews.}}]{Qiang2016}%
  \BibitemOpen
  \bibfield  {author} {\bibinfo {author} {\bibfnamefont {Xiaogang}\
  \bibnamefont {Qiang}}, \bibinfo {author} {\bibfnamefont {Thomas}\
  \bibnamefont {Loke}}, \bibinfo {author} {\bibfnamefont {Ashley}\ \bibnamefont
  {Montanaro}}, \bibinfo {author} {\bibfnamefont {Kanin}\ \bibnamefont
  {Aungskunsiri}}, \bibinfo {author} {\bibfnamefont {Xiaoqi}\ \bibnamefont
  {Zhou}}, \bibinfo {author} {\bibfnamefont {Jeremy~L.}\ \bibnamefont
  {O'Brien}}, \bibinfo {author} {\bibfnamefont {Jingbo~B.}\ \bibnamefont
  {Wang}}, \ and\ \bibinfo {author} {\bibfnamefont {Jonathan C.~F.}\
  \bibnamefont {Matthews.}},\ }\bibfield  {title} {\enquote {\bibinfo {title}
  {Efficient quantum walk on a quantum processor},}\ }\href {\doibase
  10.1038/ncomms11511} {\bibfield  {journal} {\bibinfo  {journal} {Nature
  Communications}\ } (\bibinfo {year} {2016}),\
  10.1038/ncomms11511}\BibitemShut {NoStop}%
\bibitem [{\citenamefont {Ryan}\ \emph {et~al.}(2005)\citenamefont {Ryan},
  \citenamefont {Laforest}, \citenamefont {Boileau},\ and\ \citenamefont
  {Laflamme}}]{Ryan2005}%
  \BibitemOpen
  \bibfield  {author} {\bibinfo {author} {\bibfnamefont {C.~A.}\ \bibnamefont
  {Ryan}}, \bibinfo {author} {\bibfnamefont {M.}~\bibnamefont {Laforest}},
  \bibinfo {author} {\bibfnamefont {J.~C.}\ \bibnamefont {Boileau}}, \ and\
  \bibinfo {author} {\bibfnamefont {R.}~\bibnamefont {Laflamme}},\ }\bibfield
  {title} {\enquote {\bibinfo {title} {Experimental implementation of a
  discrete-time quantum random walk on an {NMR} quantum-information
  processor},}\ }\href {\doibase 10.1103/PhysRevA.72.062317} {\bibfield
  {journal} {\bibinfo  {journal} {Phys. Rev. A}\ }\textbf {\bibinfo {volume}
  {72}},\ \bibinfo {pages} {062317} (\bibinfo {year} {2005})}\BibitemShut
  {NoStop}%
\bibitem [{\citenamefont {Lu}\ \emph {et~al.}(2010)\citenamefont {Lu},
  \citenamefont {Zhu}, \citenamefont {Zou}, \citenamefont {Peng}, \citenamefont
  {Yu}, \citenamefont {Zhang}, \citenamefont {Chen},\ and\ \citenamefont
  {Du}}]{Lu2010}%
  \BibitemOpen
  \bibfield  {author} {\bibinfo {author} {\bibfnamefont {Dawei}\ \bibnamefont
  {Lu}}, \bibinfo {author} {\bibfnamefont {Jing}\ \bibnamefont {Zhu}}, \bibinfo
  {author} {\bibfnamefont {Ping}\ \bibnamefont {Zou}}, \bibinfo {author}
  {\bibfnamefont {Xinhua}\ \bibnamefont {Peng}}, \bibinfo {author}
  {\bibfnamefont {Yihua}\ \bibnamefont {Yu}}, \bibinfo {author} {\bibfnamefont
  {Shanmin}\ \bibnamefont {Zhang}}, \bibinfo {author} {\bibfnamefont {Qun}\
  \bibnamefont {Chen}}, \ and\ \bibinfo {author} {\bibfnamefont {Jiangfeng}\
  \bibnamefont {Du}},\ }\bibfield  {title} {\enquote {\bibinfo {title}
  {Experimental implementation of a quantum random-walk search algorithm using
  strongly dipolar coupled spins},}\ }\href {\doibase
  10.1103/PhysRevA.81.022308} {\bibfield  {journal} {\bibinfo  {journal} {Phys.
  Rev. A}\ }\textbf {\bibinfo {volume} {81}},\ \bibinfo {pages} {022308}
  (\bibinfo {year} {2010})}\BibitemShut {NoStop}%
\bibitem [{\citenamefont {Matjeschk}\ \emph {et~al.}(2012)\citenamefont
  {Matjeschk}, \citenamefont {Schneider}, \citenamefont {Enderlein},
  \citenamefont {Huber}, \citenamefont {Schmitz}, \citenamefont {Glueckert},\
  and\ \citenamefont {Schaetz}}]{Matjeschk2012}%
  \BibitemOpen
  \bibfield  {author} {\bibinfo {author} {\bibfnamefont {R.}~\bibnamefont
  {Matjeschk}}, \bibinfo {author} {\bibfnamefont {Ch.}\ \bibnamefont
  {Schneider}}, \bibinfo {author} {\bibfnamefont {M.}~\bibnamefont
  {Enderlein}}, \bibinfo {author} {\bibfnamefont {T.}~\bibnamefont {Huber}},
  \bibinfo {author} {\bibfnamefont {H.}~\bibnamefont {Schmitz}}, \bibinfo
  {author} {\bibfnamefont {J.}~\bibnamefont {Glueckert}}, \ and\ \bibinfo
  {author} {\bibfnamefont {T}~\bibnamefont {Schaetz}},\ }\bibfield  {title}
  {\enquote {\bibinfo {title} {Experimental simulation and limitations of
  quantum walks with trapped ions},}\ }\href {\doibase
  10.1088/1367-2630/14/3/035012} {\bibfield  {journal} {\bibinfo  {journal}
  {New Journal of Physics}\ }\textbf {\bibinfo {volume} {14}} (\bibinfo {year}
  {2012}),\ 10.1088/1367-2630/14/3/035012}\BibitemShut {NoStop}%
\bibitem [{\citenamefont {Liber}\ and\ \citenamefont
  {Nita}(2019)}]{Arkadiusz19a}%
  \BibitemOpen
  \bibfield  {author} {\bibinfo {author} {\bibfnamefont {Arkadiusz}\
  \bibnamefont {Liber}}\ and\ \bibinfo {author} {\bibfnamefont {Laurentiu}\
  \bibnamefont {Nita}},\ }\bibfield  {title} {\enquote {\bibinfo {title} {The
  research of grover's quantum search algorithm with use of quantum circuits
  {QX2} and {QX4}},}\ }in\ \href@noop {} {\emph {\bibinfo {booktitle}
  {Information Systems Architecture and Technology: Proceedings of 39th
  International Conference on Information Systems Architecture and Technology
  -- ISAT 2018}}},\ \bibinfo {editor} {edited by\ \bibinfo {editor}
  {\bibfnamefont {Leszek}\ \bibnamefont {Borzemski}}, \bibinfo {editor}
  {\bibfnamefont {Jerzy}\ \bibnamefont {{\'{S}}wi{\k{a}}tek}}, \ and\ \bibinfo
  {editor} {\bibfnamefont {Zofia}\ \bibnamefont {Wilimowska}}}\ (\bibinfo
  {publisher} {Springer International Publishing},\ \bibinfo {address} {Cham},\
  \bibinfo {year} {2019})\ pp.\ \bibinfo {pages} {146--155}\BibitemShut
  {NoStop}%
\bibitem [{\citenamefont {Chancellor}\ \emph {et~al.}(2017)\citenamefont
  {Chancellor}, \citenamefont {Zohren},\ and\ \citenamefont
  {Warburton}}]{chancellor17a}%
  \BibitemOpen
  \bibfield  {author} {\bibinfo {author} {\bibfnamefont {Nicholas}\
  \bibnamefont {Chancellor}}, \bibinfo {author} {\bibfnamefont {Stefan}\
  \bibnamefont {Zohren}}, \ and\ \bibinfo {author} {\bibfnamefont {Paul~A.}\
  \bibnamefont {Warburton}},\ }\bibfield  {title} {\enquote {\bibinfo {title}
  {Circuit design for multi-body interactions in superconducting quantum
  annealing systems with applications to a scalable architecture},}\ }\href
  {\doibase 10.1038/s41534-017-0022-6} {\bibfield  {journal} {\bibinfo
  {journal} {npj Quantum Information}\ }\textbf {\bibinfo {volume} {3}}
  (\bibinfo {year} {2017}),\ 10.1038/s41534-017-0022-6},\ \Eprint
  {http://arxiv.org/abs/arXiv:1603.09521} {arXiv:1603.09521} \BibitemShut
  {NoStop}%
\bibitem [{\citenamefont {Farhi}\ \emph {et~al.}(2002)\citenamefont {Farhi},
  \citenamefont {Goldstone},\ and\ \citenamefont {Gutmann}}]{farhi02a}%
  \BibitemOpen
  \bibfield  {author} {\bibinfo {author} {\bibfnamefont {Edward}\ \bibnamefont
  {Farhi}}, \bibinfo {author} {\bibfnamefont {Jeffrey}\ \bibnamefont
  {Goldstone}}, \ and\ \bibinfo {author} {\bibfnamefont {Sam}\ \bibnamefont
  {Gutmann}},\ }\href@noop {} {\enquote {\bibinfo {title} {Quantum adiabatic
  evolution algorithms versus simulated annealing},}\ } (\bibinfo {year}
  {2002}),\ \Eprint {http://arxiv.org/abs/ar$\chi$iv:quant-ph/0201031}
  {ar$\chi$iv:quant-ph/0201031} \BibitemShut {NoStop}%
\bibitem [{\citenamefont {Kong}\ and\ \citenamefont {Crosson}(2015)}]{kong17a}%
  \BibitemOpen
  \bibfield  {author} {\bibinfo {author} {\bibfnamefont {L.}~\bibnamefont
  {Kong}}\ and\ \bibinfo {author} {\bibfnamefont {E.}~\bibnamefont {Crosson}},\
  }\href {\doibase 10.1142/S0219749917500113} {\enquote {\bibinfo {title} {The
  performance of the quantum adiabatic algorithm on spike hamiltonians},}\ }
  (\bibinfo {year} {2015}),\ \Eprint
  {http://arxiv.org/abs/ar$\chi$iv:1511.06991} {ar$\chi$iv:1511.06991}
  \BibitemShut {NoStop}%
\bibitem [{\citenamefont {Reichardt}(2004)}]{reichardt04a}%
  \BibitemOpen
  \bibfield  {author} {\bibinfo {author} {\bibfnamefont {B.~W.}\ \bibnamefont
  {Reichardt}},\ }\href@noop {} {\bibfield  {journal} {\bibinfo  {journal}
  {Proceedings of the Thirty Sixth Annual Symposium on the Theory of Computing
  STOC}\ ,\ \bibinfo {pages} {502--510}} (\bibinfo {year} {2004})},\ \bibinfo
  {note} {errata:
  \href{http://www-bcf.usc.edu/~breichar/Correction.txt}{http://www-bcf.usc.edu/~breichar/Correction.txt}}\BibitemShut
  {NoStop}%
\bibitem [{\citenamefont {Albash}\ and\ \citenamefont
  {Lidar}(2018)}]{albash16a}%
  \BibitemOpen
  \bibfield  {author} {\bibinfo {author} {\bibfnamefont {Tameem}\ \bibnamefont
  {Albash}}\ and\ \bibinfo {author} {\bibfnamefont {Daniel~A.}\ \bibnamefont
  {Lidar}},\ }\bibfield  {title} {\enquote {\bibinfo {title} {Adiabatic quantum
  computing},}\ }\href {\doibase 10.1103/RevModPhys.90.015002} {\bibfield
  {journal} {\bibinfo  {journal} {Rev. Mod. Phys.}\ }\textbf {\bibinfo {volume}
  {90}} (\bibinfo {year} {2018}),\ 10.1103/RevModPhys.90.015002}\BibitemShut
  {NoStop}%
\bibitem [{\citenamefont {Crosson}\ and\ \citenamefont
  {Deng}(2014)}]{Crosson14a}%
  \BibitemOpen
  \bibfield  {author} {\bibinfo {author} {\bibfnamefont {Elizabeth}\
  \bibnamefont {Crosson}}\ and\ \bibinfo {author} {\bibfnamefont {Mingkai}\
  \bibnamefont {Deng}},\ }\href@noop {} {\enquote {\bibinfo {title} {Tunneling
  through high energy barriers in simulated quantum annealing},}\ } (\bibinfo
  {year} {2014}),\ \Eprint {http://arxiv.org/abs/arXiv:1410.8484}
  {arXiv:1410.8484} \BibitemShut {NoStop}%
\bibitem [{\citenamefont {Brady}\ and\ \citenamefont {van
  Dam}(2016)}]{Brady15a}%
  \BibitemOpen
  \bibfield  {author} {\bibinfo {author} {\bibfnamefont {Lucas~T.}\
  \bibnamefont {Brady}}\ and\ \bibinfo {author} {\bibfnamefont {Wim}\
  \bibnamefont {van Dam}},\ }\bibfield  {title} {\enquote {\bibinfo {title}
  {Quantum monte carlo simulations of tunneling in quantum adiabatic
  optimization},}\ }\href {\doibase 10.1103/PhysRevA.93.032304} {\bibfield
  {journal} {\bibinfo  {journal} {Phys. Rev. A}\ }\textbf {\bibinfo {volume}
  {93}} (\bibinfo {year} {2016}),\ 10.1103/PhysRevA.93.032304},\ \Eprint
  {http://arxiv.org/abs/arXiv:1509.02562} {arXiv:1509.02562} \BibitemShut
  {NoStop}%
\bibitem [{\citenamefont {Crosson}\ and\ \citenamefont
  {Harrow}(2016)}]{Crosson16a}%
  \BibitemOpen
  \bibfield  {author} {\bibinfo {author} {\bibfnamefont {Elizabeth}\
  \bibnamefont {Crosson}}\ and\ \bibinfo {author} {\bibfnamefont {Aram~W.}\
  \bibnamefont {Harrow}},\ }\bibfield  {title} {\enquote {\bibinfo {title}
  {Simulated quantum annealing can be exponentially faster than classical
  simulated annealing},}\ }\href {\doibase 10.1109/FOCS.2016.81} {\bibfield
  {journal} {\bibinfo  {journal} {Proc of FOCS 2016, pp. 714-723}\ } (\bibinfo
  {year} {2016}),\ 10.1109/FOCS.2016.81},\ \Eprint
  {http://arxiv.org/abs/arXiv:1601.03030} {arXiv:1601.03030} \BibitemShut
  {NoStop}%
\bibitem [{\citenamefont {Muthukrishnan}\ \emph {et~al.}(2016)\citenamefont
  {Muthukrishnan}, \citenamefont {Albash},\ and\ \citenamefont
  {Lidar}}]{muthukrishnan16a}%
  \BibitemOpen
  \bibfield  {author} {\bibinfo {author} {\bibfnamefont {Siddharth}\
  \bibnamefont {Muthukrishnan}}, \bibinfo {author} {\bibfnamefont {Tameem}\
  \bibnamefont {Albash}}, \ and\ \bibinfo {author} {\bibfnamefont {Daniel~A.}\
  \bibnamefont {Lidar}},\ }\bibfield  {title} {\enquote {\bibinfo {title}
  {Tunneling and speedup in quantum optimization for permutation-symmetric
  problems},}\ }\href {\doibase 10.1103/PhysRevX.6.031010} {\bibfield
  {journal} {\bibinfo  {journal} {Phys. Rev. X}\ }\textbf {\bibinfo {volume}
  {6}},\ \bibinfo {pages} {031010} (\bibinfo {year} {2016})}\BibitemShut
  {NoStop}%
\bibitem [{\citenamefont {Brady}\ and\ \citenamefont {van
  Dam}(2017)}]{Brady17a}%
  \BibitemOpen
  \bibfield  {author} {\bibinfo {author} {\bibfnamefont {Lucas~T.}\
  \bibnamefont {Brady}}\ and\ \bibinfo {author} {\bibfnamefont {Wim}\
  \bibnamefont {van Dam}},\ }\bibfield  {title} {\enquote {\bibinfo {title}
  {Necessary adiabatic run times in quantum optimization},}\ }\href {\doibase
  10.1103/PhysRevA.95.032335} {\bibfield  {journal} {\bibinfo  {journal} {Phys.
  Rev. A}\ }\textbf {\bibinfo {volume} {95}} (\bibinfo {year} {2017}),\
  10.1103/PhysRevA.95.032335},\ \Eprint {http://arxiv.org/abs/arXiv:1611.02585}
  {arXiv:1611.02585} \BibitemShut {NoStop}%
\bibitem [{\citenamefont {Fulton}\ \emph {et~al.}(2019)\citenamefont {Fulton},
  \citenamefont {Wilson}, \citenamefont {Kulmiya}, \citenamefont {Dodds},
  \citenamefont {Chancellor},\ and\ \citenamefont {Kendon}}]{fulton18a}%
  \BibitemOpen
  \bibfield  {author} {\bibinfo {author} {\bibfnamefont {James}\ \bibnamefont
  {Fulton}}, \bibinfo {author} {\bibfnamefont {Zac}\ \bibnamefont {Wilson}},
  \bibinfo {author} {\bibfnamefont {Sahra}\ \bibnamefont {Kulmiya}}, \bibinfo
  {author} {\bibfnamefont {Ben}\ \bibnamefont {Dodds}}, \bibinfo {author}
  {\bibfnamefont {Nicholas}\ \bibnamefont {Chancellor}}, \ and\ \bibinfo
  {author} {\bibfnamefont {Viv}\ \bibnamefont {Kendon}},\ }\href@noop {}
  {\enquote {\bibinfo {title} {Permutation symmetric quantum computing:
  gadgets, adaiatic quantum computation, and quantum walks},}\ } (\bibinfo
  {year} {2019}),\ \bibinfo {note} {in preparation}\BibitemShut {NoStop}%
\bibitem [{\citenamefont {Dattani}(2019)}]{Dattani18a}%
  \BibitemOpen
  \bibfield  {author} {\bibinfo {author} {\bibfnamefont {Nikesh~S.}\
  \bibnamefont {Dattani}},\ }\href@noop {} {\enquote {\bibinfo {title}
  {Quadratization in discrete optimization and quantum mechanics},}\ }
  (\bibinfo {year} {2019}),\ \Eprint {http://arxiv.org/abs/1901.04405}
  {arXiv:1901.04405} \BibitemShut {NoStop}%
\bibitem [{\citenamefont {Leib}\ \emph {et~al.}(2016)\citenamefont {Leib},
  \citenamefont {Zoller},\ and\ \citenamefont {Lechner}}]{Leib16a}%
  \BibitemOpen
  \bibfield  {author} {\bibinfo {author} {\bibfnamefont {Martin}\ \bibnamefont
  {Leib}}, \bibinfo {author} {\bibfnamefont {Peter}\ \bibnamefont {Zoller}}, \
  and\ \bibinfo {author} {\bibfnamefont {Wolfgang}\ \bibnamefont {Lechner}},\
  }\bibfield  {title} {\enquote {\bibinfo {title} {A transmon quantum annealer:
  decomposing many-body ising constraints into pair interactions},}\ }\href
  {http://stacks.iop.org/2058-9565/1/i=1/a=015008} {\bibfield  {journal}
  {\bibinfo  {journal} {Quantum Science and Technology}\ }\textbf {\bibinfo
  {volume} {1}},\ \bibinfo {pages} {015008} (\bibinfo {year}
  {2016})}\BibitemShut {NoStop}%
\bibitem [{\citenamefont {Jordan}\ and\ \citenamefont
  {Farhi}(2008)}]{Jordan08a}%
  \BibitemOpen
  \bibfield  {author} {\bibinfo {author} {\bibfnamefont {Stephen~P.}\
  \bibnamefont {Jordan}}\ and\ \bibinfo {author} {\bibfnamefont {Edward}\
  \bibnamefont {Farhi}},\ }\bibfield  {title} {\enquote {\bibinfo {title}
  {Perturbative gadgets at arbitrary orders},}\ }\href {\doibase
  10.1103/PhysRevA.77.062329} {\bibfield  {journal} {\bibinfo  {journal} {Phys.
  Rev. A 77, 062329 (2008)}\ } (\bibinfo {year} {2008}),\
  10.1103/PhysRevA.77.062329},\ \Eprint {http://arxiv.org/abs/arXiv:0802.1874}
  {arXiv:0802.1874} \BibitemShut {NoStop}%
\bibitem [{\citenamefont {Kempe}\ \emph {et~al.}(2006)\citenamefont {Kempe},
  \citenamefont {Kitaev},\ and\ \citenamefont {Regev}}]{Kempe04a}%
  \BibitemOpen
  \bibfield  {author} {\bibinfo {author} {\bibfnamefont {Julia}\ \bibnamefont
  {Kempe}}, \bibinfo {author} {\bibfnamefont {Alexei}\ \bibnamefont {Kitaev}},
  \ and\ \bibinfo {author} {\bibfnamefont {Oded}\ \bibnamefont {Regev}},\
  }\bibfield  {title} {\enquote {\bibinfo {title} {The complexity of the local
  hamiltonian problem},}\ }\href@noop {} {\bibfield  {journal} {\bibinfo
  {journal} {SIAM Journal of Computing, conference version in Proc. 24th
  FSTTCS, p. 372-383 (2004)}\ }\textbf {\bibinfo {volume} {35(5)}},\ \bibinfo
  {pages} {1070--1097} (\bibinfo {year} {2006})},\ \Eprint
  {http://arxiv.org/abs/arXiv:quant-ph/0406180} {arXiv:quant-ph/0406180}
  \BibitemShut {NoStop}%
\bibitem [{\citenamefont {Biamonte}\ and\ \citenamefont
  {Love}(2007)}]{Biamonte07a}%
  \BibitemOpen
  \bibfield  {author} {\bibinfo {author} {\bibfnamefont {Jacob~D.}\
  \bibnamefont {Biamonte}}\ and\ \bibinfo {author} {\bibfnamefont {Peter~J.}\
  \bibnamefont {Love}},\ }\bibfield  {title} {\enquote {\bibinfo {title}
  {Realizable hamiltonians for universal adiabatic quantum computers},}\ }\href
  {\doibase 10.1103/PhysRevA.78.012352} {\bibfield  {journal} {\bibinfo
  {journal} {Phys. Rev. A 78, 012352 (2008).}\ } (\bibinfo {year} {2007}),\
  10.1103/PhysRevA.78.012352},\ \Eprint {http://arxiv.org/abs/arXiv:0704.1287}
  {arXiv:0704.1287} \BibitemShut {NoStop}%
\bibitem [{\citenamefont {Le~Bellac}(2006)}]{LeBellac}%
  \BibitemOpen
  \bibfield  {author} {\bibinfo {author} {\bibfnamefont {Michel}\ \bibnamefont
  {Le~Bellac}},\ }\href@noop {} {\emph {\bibinfo {title} {Quantum Physics}}}\
  (\bibinfo  {publisher} {Cambridge University Press},\ \bibinfo {year}
  {2006})\BibitemShut {NoStop}%
\bibitem [{\citenamefont {Saffman}(2016)}]{Saffman16a}%
  \BibitemOpen
  \bibfield  {author} {\bibinfo {author} {\bibfnamefont {M}~\bibnamefont
  {Saffman}},\ }\bibfield  {title} {\enquote {\bibinfo {title} {Quantum
  computing with atomic qubits and rydberg interactions: progress and
  challenges},}\ }\href {http://stacks.iop.org/0953-4075/49/i=20/a=202001}
  {\bibfield  {journal} {\bibinfo  {journal} {Journal of Physics B: Atomic,
  Molecular and Optical Physics}\ }\textbf {\bibinfo {volume} {49}},\ \bibinfo
  {pages} {202001} (\bibinfo {year} {2016})}\BibitemShut {NoStop}%
\bibitem [{\citenamefont {\v{S}ibali\'c}\ and\ \citenamefont
  {Adams}(2018)}]{Sibalic18a}%
  \BibitemOpen
  \bibfield  {author} {\bibinfo {author} {\bibfnamefont {Nikola}\ \bibnamefont
  {\v{S}ibali\'c}}\ and\ \bibinfo {author} {\bibfnamefont {Charles~S}\
  \bibnamefont {Adams}},\ }\bibfield  {title} {\enquote {\bibinfo {title}
  {Rydberg physics},}\ }in\ \href {\doibase 10.1088/978-0-7503-1635-4ch1}
  {\emph {\bibinfo {booktitle} {Rydberg Physics}}},\ \bibinfo {series and
  number} {2399-2891}\ (\bibinfo  {publisher} {IOP Publishing},\ \bibinfo
  {year} {2018})\ pp.\ \bibinfo {pages} {1--1 to 1--27}\BibitemShut {NoStop}%
\bibitem [{\citenamefont {Barredo}\ \emph {et~al.}(2018)\citenamefont
  {Barredo}, \citenamefont {Lienhard}, \citenamefont {de~L{\'e}s{\'e}leuc},
  \citenamefont {Lahaye},\ and\ \citenamefont {Browaeys}}]{Barredo18a}%
  \BibitemOpen
  \bibfield  {author} {\bibinfo {author} {\bibfnamefont {Daniel}\ \bibnamefont
  {Barredo}}, \bibinfo {author} {\bibfnamefont {Vincent}\ \bibnamefont
  {Lienhard}}, \bibinfo {author} {\bibfnamefont {Sylvain}\ \bibnamefont
  {de~L{\'e}s{\'e}leuc}}, \bibinfo {author} {\bibfnamefont {Thierry}\
  \bibnamefont {Lahaye}}, \ and\ \bibinfo {author} {\bibfnamefont {Antoine}\
  \bibnamefont {Browaeys}},\ }\bibfield  {title} {\enquote {\bibinfo {title}
  {Synthetic three-dimensional atomic structures assembled atom by atom},}\
  }\href {\doibase 10.1038/s41586-018-0450-2} {\bibfield  {journal} {\bibinfo
  {journal} {Nature}\ }\textbf {\bibinfo {volume} {561}},\ \bibinfo {pages}
  {79--82} (\bibinfo {year} {2018})}\BibitemShut {NoStop}%
\bibitem [{\citenamefont {Levine}\ \emph {et~al.}(2018)\citenamefont {Levine},
  \citenamefont {Keesling}, \citenamefont {Omran}, \citenamefont {Bernien},
  \citenamefont {Schwartz}, \citenamefont {Zibrov}, \citenamefont {Endres},
  \citenamefont {Greiner}, \citenamefont {Vuleti\'{c}},\ and\ \citenamefont
  {Lukin}}]{Levine18a}%
  \BibitemOpen
  \bibfield  {author} {\bibinfo {author} {\bibfnamefont {Harry}\ \bibnamefont
  {Levine}}, \bibinfo {author} {\bibfnamefont {Alexander}\ \bibnamefont
  {Keesling}}, \bibinfo {author} {\bibfnamefont {Ahmed}\ \bibnamefont {Omran}},
  \bibinfo {author} {\bibfnamefont {Hannes}\ \bibnamefont {Bernien}}, \bibinfo
  {author} {\bibfnamefont {Sylvain}\ \bibnamefont {Schwartz}}, \bibinfo
  {author} {\bibfnamefont {Alexander~S.}\ \bibnamefont {Zibrov}}, \bibinfo
  {author} {\bibfnamefont {Manuel}\ \bibnamefont {Endres}}, \bibinfo {author}
  {\bibfnamefont {Markus}\ \bibnamefont {Greiner}}, \bibinfo {author}
  {\bibfnamefont {Vladan}\ \bibnamefont {Vuleti\'{c}}}, \ and\ \bibinfo
  {author} {\bibfnamefont {Mikhail~D.}\ \bibnamefont {Lukin}},\ }\bibfield
  {title} {\enquote {\bibinfo {title} {High-fidelity control and entanglement
  of rydberg-atom qubits},}\ }\href {\doibase 10.1103/PhysRevLett.121.123603}
  {\bibfield  {journal} {\bibinfo  {journal} {Phys. Rev. Lett.}\ }\textbf
  {\bibinfo {volume} {121}},\ \bibinfo {pages} {123603} (\bibinfo {year}
  {2018})}\BibitemShut {NoStop}%
\bibitem [{\citenamefont {Picken}\ \emph {et~al.}(2018)\citenamefont {Picken},
  \citenamefont {Legaie}, \citenamefont {McDonnell},\ and\ \citenamefont
  {Pritchard}}]{Picken18a}%
  \BibitemOpen
  \bibfield  {author} {\bibinfo {author} {\bibfnamefont {C~J}\ \bibnamefont
  {Picken}}, \bibinfo {author} {\bibfnamefont {R}~\bibnamefont {Legaie}},
  \bibinfo {author} {\bibfnamefont {K}~\bibnamefont {McDonnell}}, \ and\
  \bibinfo {author} {\bibfnamefont {J~D}\ \bibnamefont {Pritchard}},\
  }\bibfield  {title} {\enquote {\bibinfo {title} {Entanglement of neutral-atom
  qubits with long ground-rydberg coherence times},}\ }\href {\doibase
  10.1088/2058-9565/aaf019} {\bibfield  {journal} {\bibinfo  {journal} {Quantum
  Science and Technology}\ }\textbf {\bibinfo {volume} {4}},\ \bibinfo {pages}
  {015011} (\bibinfo {year} {2018})}\BibitemShut {NoStop}%
\bibitem [{\citenamefont {Cooper}\ \emph {et~al.}(2018)\citenamefont {Cooper},
  \citenamefont {Covey}, \citenamefont {Madjarov}, \citenamefont {Porsev},
  \citenamefont {Safronova},\ and\ \citenamefont {Endres}}]{Cooper18a}%
  \BibitemOpen
  \bibfield  {author} {\bibinfo {author} {\bibfnamefont {A.}~\bibnamefont
  {Cooper}}, \bibinfo {author} {\bibfnamefont {J.~P.}\ \bibnamefont {Covey}},
  \bibinfo {author} {\bibfnamefont {I.~S.}\ \bibnamefont {Madjarov}}, \bibinfo
  {author} {\bibfnamefont {S.~G.}\ \bibnamefont {Porsev}}, \bibinfo {author}
  {\bibfnamefont {M.~S.}\ \bibnamefont {Safronova}}, \ and\ \bibinfo {author}
  {\bibfnamefont {M.}~\bibnamefont {Endres}},\ }\bibfield  {title} {\enquote
  {\bibinfo {title} {Alkaline-earth atoms in optical tweezers},}\ }\href
  {\doibase 10.1103/PhysRevX.8.041055} {\bibfield  {journal} {\bibinfo
  {journal} {Phys.~Rev.~X}\ }\textbf {\bibinfo {volume} {8}},\ \bibinfo {pages}
  {041055} (\bibinfo {year} {2018})}\BibitemShut {NoStop}%
\bibitem [{\citenamefont {M.~A.~Norcia}\ and\ \citenamefont
  {Kaufman}(2018)}]{Norcia18a}%
  \BibitemOpen
  \bibfield  {author} {\bibinfo {author} {\bibfnamefont {A.~W.~Young}\
  \bibnamefont {M.~A.~Norcia}}\ and\ \bibinfo {author} {\bibfnamefont {A.~M.}\
  \bibnamefont {Kaufman}},\ }\bibfield  {title} {\enquote {\bibinfo {title}
  {Microscopic control and detection of ultracold strontium in optical-tweezer
  arrays},}\ }\href {\doibase 10.1103/PhysRevX.8.041054} {\bibfield  {journal}
  {\bibinfo  {journal} {Phys.~Rev.~X}\ }\textbf {\bibinfo {volume} {8}},\
  \bibinfo {pages} {041054} (\bibinfo {year} {2018})}\BibitemShut {NoStop}%
\bibitem [{\citenamefont {Saskin}\ \emph {et~al.}(2018)\citenamefont {Saskin},
  \citenamefont {Wilson}, \citenamefont {Grinkemeyer},\ and\ \citenamefont
  {Thompson}}]{Saskin18a}%
  \BibitemOpen
  \bibfield  {author} {\bibinfo {author} {\bibfnamefont {S.}~\bibnamefont
  {Saskin}}, \bibinfo {author} {\bibfnamefont {J.}~\bibnamefont {Wilson}},
  \bibinfo {author} {\bibfnamefont {B.}~\bibnamefont {Grinkemeyer}}, \ and\
  \bibinfo {author} {\bibfnamefont {J.}~\bibnamefont {Thompson}},\ }\href@noop
  {} {\enquote {\bibinfo {title} {Narrow-line cooling and imaging of ytterbium
  atoms in an optical tweezer array},}\ } (\bibinfo {year} {2018}),\ \Eprint
  {http://arxiv.org/abs/arXiv:1810.10517} {arXiv:1810.10517} \BibitemShut
  {NoStop}%
\bibitem [{\citenamefont {Glover}\ and\ \citenamefont
  {Kochenberger}(2018)}]{Glover18a}%
  \BibitemOpen
  \bibfield  {author} {\bibinfo {author} {\bibfnamefont {Fred}\ \bibnamefont
  {Glover}}\ and\ \bibinfo {author} {\bibfnamefont {Gary}\ \bibnamefont
  {Kochenberger}},\ }\href@noop {} {\enquote {\bibinfo {title} {A tutorial on
  formulating and using qubo models},}\ } (\bibinfo {year} {2018}),\ \Eprint
  {http://arxiv.org/abs/ar$\chi$iv:1811.11538} {ar$\chi$iv:1811.11538}
  \BibitemShut {NoStop}%
\bibitem [{\citenamefont {Lucas}(2014)}]{Lucas14a}%
  \BibitemOpen
  \bibfield  {author} {\bibinfo {author} {\bibfnamefont {Andrew}\ \bibnamefont
  {Lucas}},\ }\bibfield  {title} {\enquote {\bibinfo {title} {Ising
  formulations of many np problems},}\ }\href {\doibase
  10.3389/fphy.2014.00005} {\bibfield  {journal} {\bibinfo  {journal}
  {Frontiers in Physics}\ }\textbf {\bibinfo {volume} {2}},\ \bibinfo {pages}
  {5} (\bibinfo {year} {2014})}\BibitemShut {NoStop}%
\bibitem [{\citenamefont {Bounds}\ \emph {et~al.}(2018)\citenamefont {Bounds},
  \citenamefont {Jackson}, \citenamefont {Hanley}, \citenamefont {Faoro},
  \citenamefont {Bridge}, \citenamefont {Huillery},\ and\ \citenamefont
  {Jones}}]{Bounds18a}%
  \BibitemOpen
  \bibfield  {author} {\bibinfo {author} {\bibfnamefont {A.~D.}\ \bibnamefont
  {Bounds}}, \bibinfo {author} {\bibfnamefont {N.~C.}\ \bibnamefont {Jackson}},
  \bibinfo {author} {\bibfnamefont {R.~K.}\ \bibnamefont {Hanley}}, \bibinfo
  {author} {\bibfnamefont {R.}~\bibnamefont {Faoro}}, \bibinfo {author}
  {\bibfnamefont {E.~M.}\ \bibnamefont {Bridge}}, \bibinfo {author}
  {\bibfnamefont {P.}~\bibnamefont {Huillery}}, \ and\ \bibinfo {author}
  {\bibfnamefont {M.~P.~A.}\ \bibnamefont {Jones}},\ }\bibfield  {title}
  {\enquote {\bibinfo {title} {Rydberg-dressed magneto-optical trap},}\ }\href
  {\doibase 10.1103/PhysRevLett.120.183401} {\bibfield  {journal} {\bibinfo
  {journal} {Phys. Rev. Lett.}\ }\textbf {\bibinfo {volume} {120}},\ \bibinfo
  {pages} {183401} (\bibinfo {year} {2018})}\BibitemShut {NoStop}%
\bibitem [{\citenamefont {de~L\'es\'eleuc}\ \emph {et~al.}(2018)\citenamefont
  {de~L\'es\'eleuc}, \citenamefont {Lienhard}, \citenamefont {Scholl},
  \citenamefont {Barredo}, \citenamefont {Weber}, \citenamefont {Lang},
  \citenamefont {B\"uchler}, \citenamefont {Lahaye},\ and\ \citenamefont
  {Browaeys}}]{Leseleuc18a}%
  \BibitemOpen
  \bibfield  {author} {\bibinfo {author} {\bibfnamefont {Sylvain}\ \bibnamefont
  {de~L\'es\'eleuc}}, \bibinfo {author} {\bibfnamefont {Vincent}\ \bibnamefont
  {Lienhard}}, \bibinfo {author} {\bibfnamefont {Pascal}\ \bibnamefont
  {Scholl}}, \bibinfo {author} {\bibfnamefont {Daniel}\ \bibnamefont
  {Barredo}}, \bibinfo {author} {\bibfnamefont {Sebastian}\ \bibnamefont
  {Weber}}, \bibinfo {author} {\bibfnamefont {Nicolai}\ \bibnamefont {Lang}},
  \bibinfo {author} {\bibfnamefont {Hans~Peter}\ \bibnamefont {B\"uchler}},
  \bibinfo {author} {\bibfnamefont {Thierry}\ \bibnamefont {Lahaye}}, \ and\
  \bibinfo {author} {\bibfnamefont {Antoine}\ \bibnamefont {Browaeys}},\
  }\href@noop {} {\enquote {\bibinfo {title} {Experimental realization of a
  symmetry protected topological phase of interacting bosons with rydberg
  atoms},}\ } (\bibinfo {year} {2018}),\ \Eprint
  {http://arxiv.org/abs/arXiv:1810.13286} {arXiv:1810.13286} \BibitemShut
  {NoStop}%
\bibitem [{\citenamefont {Paredes-Barato}\ and\ \citenamefont
  {Adams}(2014)}]{Paredes-Barato14a}%
  \BibitemOpen
  \bibfield  {author} {\bibinfo {author} {\bibfnamefont {D.}~\bibnamefont
  {Paredes-Barato}}\ and\ \bibinfo {author} {\bibfnamefont {C.~S.}\
  \bibnamefont {Adams}},\ }\bibfield  {title} {\enquote {\bibinfo {title}
  {All-optical quantum information processing using rydberg gates},}\ }\href
  {\doibase 10.1103/PhysRevLett.112.040501} {\bibfield  {journal} {\bibinfo
  {journal} {Phys. Rev. Lett.}\ }\textbf {\bibinfo {volume} {112}},\ \bibinfo
  {pages} {040501} (\bibinfo {year} {2014})}\BibitemShut {NoStop}%
\bibitem [{tik(2013)}]{tikzit}%
  \BibitemOpen
  \bibfield  {title} {\enquote {\bibinfo {title} {Tikzit},}\ }\href@noop {}
  {\bibfield  {journal} {\bibinfo  {journal} {http://tikzit.github.io/}\ }
  (\bibinfo {year} {2013})},\ \bibinfo {note} {accessed April 20th,
  2018}\BibitemShut {NoStop}%
\bibitem [{pyt(2016)}]{python3}%
  \BibitemOpen
  \href {https://www.python.org/} {\enquote {\bibinfo {title} {{P}ython 3.5},}\
  } (\bibinfo {year} {2016}),\ \bibinfo {note} {https://www.python.org/,
  accessed August 10th, 2016}\BibitemShut {NoStop}%
\bibitem [{num(2016)}]{numpy}%
  \BibitemOpen
  \href {http://www.numpy.org/} {\enquote {\bibinfo {title} {Numpy 1.11.1},}\ }
  (\bibinfo {year} {2016}),\ \bibinfo {note} {http://www.numpy.org/, accessed
  August 10th, 2016}\BibitemShut {NoStop}%
\bibitem [{mat(2016)}]{matplotlib}%
  \BibitemOpen
  \href {http://matplotlib.org/} {\enquote {\bibinfo {title} {Matplotlib
  1.5.1},}\ } (\bibinfo {year} {2016}),\ \bibinfo {note}
  {http://matplotlib.org/, accessed August 10th, 2016}\BibitemShut {NoStop}%
\end{thebibliography}

%

\end{document}